\documentclass[12pt]{article}
\usepackage{amsmath}
\usepackage{graphicx,psfrag,epsf}
\usepackage{enumerate}
\usepackage{natbib}
\usepackage[hyphens]{url}
\usepackage{amsfonts}
\usepackage{mathrsfs}
\usepackage{multirow}
\usepackage{hhline}
\usepackage{subfigure}
\usepackage{graphicx}
\usepackage{booktabs}
\usepackage{float}

\usepackage{algorithm}
\usepackage{algorithmicx}

\usepackage{algpseudocode}
\usepackage{authblk}

\usepackage{setspace}
\usepackage{slashbox}

\usepackage{relsize}

\newcommand{\mb}[1]{\mathbf{#1}}
\newcommand{\sbf}[1]{\boldsymbol{#1}}

\newcommand{\blind}{0}

\begin{document}

\def\spacingset#1{\renewcommand{\baselinestretch}%
{#1}\small\normalsize} \spacingset{1}

\spacingset{1.2}

\if0\blind
{
  \title{\LARGE\bf Bayesian Inference of Spatio-Temporal Changes of Arctic Sea Ice}
  \author{Bohai Zhang  \\
School of Statistics and Data Science, KLMDASR, and LPMC \\ 
Nankai University, Tianjin, 300071, China  \\ 
email: \texttt{bohaizhang@nankai.edu.cn} \\
and\\
Noel Cressie\\
National Institute for Applied Statistics Research Australia,\\
University of Wollongong, NSW, 2522, Australia\\
email: \texttt{ncressie@uow.edu.au} \\}
  \date{}
  \maketitle
} \fi

\if1\blind
{
  \bigskip
  \bigskip
  \bigskip
  \begin{center}
    {\LARGE\bf Bayesian Inference of Spatio-Temporal Changes of Arctic Sea Ice}
\end{center}
  \medskip
} \fi

\bigskip
\begin{abstract}
Arctic sea ice extent has drawn increasing interest and alarm from geoscientists, owing to its rapid decline. In this article, we propose a Bayesian spatio-temporal hierarchical statistical model for binary Arctic sea ice data over two decades, where a latent dynamic spatio-temporal Gaussian process is used to model the data-dependence through a logit link function. Our ultimate goal is to perform inference on the dynamic spatial behavior of Arctic sea ice over a period of two decades. Physically motivated covariates are assessed using autologistic diagnostics. Our Bayesian spatio-temporal model shows how parameter uncertainty in such a complex hierarchical model can influence spatio-temporal prediction. The posterior distributions of new summary statistics are proposed to detect the changing patterns of Arctic sea ice over two decades since 1997.  
\end{abstract}

\noindent%
\vfill
\newpage
\spacingset{1.2} 

\section{Introduction}
Arctic sea ice extent has drawn increasing interest and alarm from geoscientists, owing to its rapid decline, particularly in the Boreal summer and early fall. This is driven by more rapid warming in the Arctic compared with other regions, a phenomenon known as the Arctic amplification \citep{cohen2014recent}. The declining sea ice directly affects the biogeochemical cycle and animals in the Arctic region such as the polar bear and seabirds \citep{meier2014arctic}. Furthermore, the changes of Arctic sea ice can lead to changing climates in other regions of the world. For example, recent studies show that the decline of Arctic sea ice can cause extreme weather in mid-latitude regions \citep[e.g., increasing the chance of cold Eurasian winters: ][]{Screen2013Exploring,cohen2014recent,mori2014robust} and influence rainfall in the state of California \citep[e.g.,][]{Cvijanovic2017Future}.
Critically, the declining sea ice will result in an albedo\textendash ice feedback effect \citep[e.g.,][]{Kumar2010Contribution,screen2013atmospheric,pistone2014observational}, where a darker Earth surface due to less sea ice leads to additional melting. 

Therefore, monitoring the spatio-temporal dynamics of Arctic sea ice is a critical component of the study of climate change. There could also be commercial interest: The loss of summer sea ice may open new shipping lanes, providing a new passage between Earth's major oceans. 

In this article, we consider Arctic sea-ice-extent data obtained from a database of Arctic sea-ice concentrations (proportion of sea ice in a grid cell) produced by the National Oceanic and Atmospheric Administration (NOAA) as part of their National Snow \& Ice Data Center's (NSIDC) Climate Data Record (CDR). The data are based on passive microwave remotely sensed data provided by the Nimbus 7 satellite and the F8, F11, and F13 satellites of the Defense Meteorological Satellite Program \citep{parkinson1999arctic,parkinson2014global}, and they are projected onto approximately $25\textnormal{km}\times25\textnormal{km}$ grid cells (or pixels). A $15\%$ cut-off is the standard used to determine whether a pixel is water ($<15\%$) or ice ($\ge15\%$), and Arctic sea ice extent (SIE) is defined as the total area of ice pixels in the Arctic region. \citep[e.g.,][]{parkinson1999arctic,zwally2002variability,meier2007whither,parkinson2014global}. 

\begin{figure}[!ht]
\centering
\includegraphics[width=12cm, height=6cm]{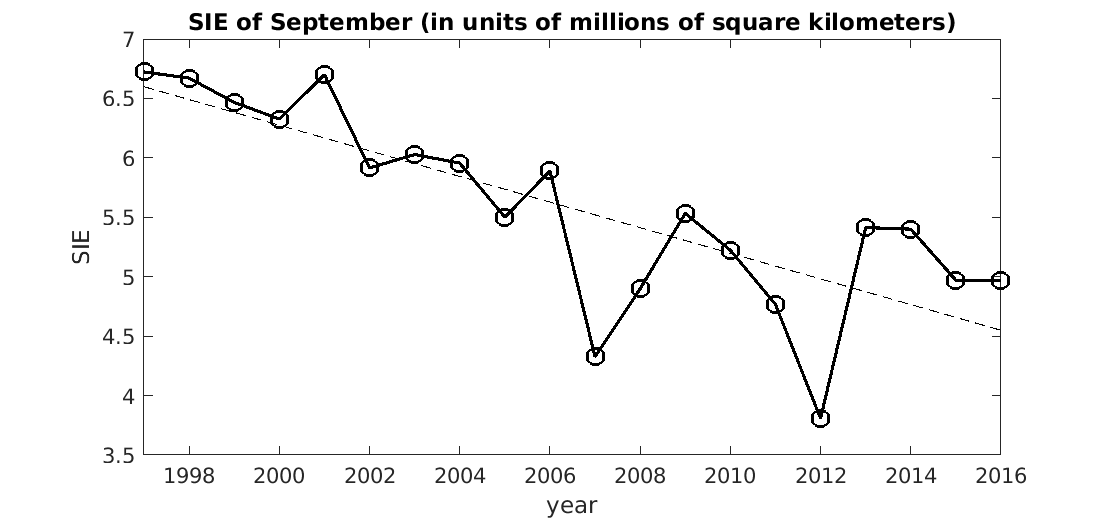}
\caption{September Arctic sea ice extent (SIE, in units of millions of km$^2$) from 1997 to 2016. Arctic SIE is the total area of ice pixels in the Arctic region. The dashed line shows an ordinary-least-squares fit. } \label{Fig_data_20years}
\end{figure}

The resulting binary ($0=\textnormal{water}$ and $1=\textnormal{ice}$) data are available monthly from November, 1978; here we focus on the month of September for the two decades 1997 \textendash~2016, since the Arctic sea-ice cover is typically least in that month of the year \citep{parkinson2014global}. Based on the September data from 1997 \textendash~2016, Figure~\ref{Fig_data_20years} shows a time series plot of the SIE in millions of km$^2$ for the Arctic region. 

Previous analyses of Arctic sea ice by geoscientists have focussed mainly on the temporal aspect or the spatial aspect, without considering both simultaneously \citep[e.g.,][]{parkinson2014global,parkinson2014spatially,parkinson2016new}. Notably, uncertainty measures for the summary statistics presented in their studies are either absent or not appropriate, given the dependence in the data. Related research on calibrating spatial binary outputs from computer models of Antarctic ice sheets can be found in \cite{chang2016calibrating,chang2016improving}. More recently, \cite{olson2019novel} proposed a new method for assessing the dependence of Arctic sea ice on climate variables (in the form of output from climate models), for the purpose of forecasting the Arctic sea ice. In contrast to these papers, the article by \cite{director2017improved,director2019probabilistic} takes a decidedly statistical approach, using a spatio-temporal statistical model to forecast selected contours of sea-ice concentration.

In this article, we use a Bayesian statistical approach for inference on Arctic sea ice based on interpretable spatial summaries for the months of September spanning two decades from 1997~\textendash~2016. Specifically, we use a spatio-temporal hierarchical generalized linear mixed model (ST-GLMM) framework, where a latent Gaussian process (GP) is introduced to model the (spatio-temporal) dependence in the data, linked to the non-Gaussian (binary) observations through a logit link. Recall that the sea-ice-extent data are defined on spatial pixels of a nominal area of $25\textnormal{km}\times 25\textnormal{km}$, and hence these pixels can be treated as Basic Areal Units (BAUs) for modeling the spatial data \citep[e.g.,][]{nguyen2012spatial, zammit2019frk}, and their centroids are used as the spatial locations of the BAU-level data. We carry out Bayesian inference on areal-based quantities for the observation period, in contrast to \cite{director2017improved,director2019probabilistic}, who forecast selected sea-ice-concentration contours beyond their observation periods. 

Our statistical model is motivated by \cite{diggle1998model}, who analyzed spatial-only binary data. When the data set is very large, which is often the case for spatio-temporal geophysical data, inference on the latent GP in a ST-GLMM is very expensive and quickly becomes prohibitive. Then the modeling strategy of using low-rank or sparse-matrix approximations can be applied to facilitate the inference. This has proved to be very effective for Gaussian spatial data \citep[e.g.,][]{furrer2006covariance,banerjee2008gaussian,cressie2008frk,rue2009approximate,Sang2012A,nychka2015multiresolution,Datta2016Hierarchical,katzfuss2017multi}. 
The approach has been adapted to analyze large non-Gaussian spatial and spatio-temporal data by \textit{inter alia} \cite{sengupta2013hierarchical}, \cite{sengupta2016predictive}, \cite{holan2015hierarchical}, \cite{bradley2016bayesian}, \cite{guan2016computationally}, \cite{bradley2018computationally}, and \cite{linero2017multi}. 

In this article, we use physical knowledge supported by data to build a scientifically motivated latent process that is Gaussian on the logistic scale. Then we put prior distributions on the ST-GLMM model parameters and use a Bayesian hierarchical model (BHM) to obtain the joint posterior distribution of parameters and the latent GP given the binary SIE data. In an earlier paper,  an empirical hierarchical modeling (EHM) approach was taken by \cite{zhang2018estimating} to analyze the binary SIE data with a single covariate given by distance to the North Pole. 

Here, the covariates that are used to model the Arctic SIE data include the averaged Arctic surface temperature anomalies in the previous summer season and in the previous winter season, in order to forecast the presence/absence of sea ice in the following September. The effects of longitude and distance to the coastline are also found to be important covariates. While generic hierarchical modeling uses Bayes' Theorem, the model becomes fully Bayesian when all parameters have prior distributions put on them. In contrast to this BHM approach, an EHM approach involves ``plugging in" the parameter estimates into the hierarchical model. 

In this article, our BHM is in fact a hybrid \citep[e.g.,][Ch.~1]{Wikle_Mangion_Cressie_2019}, where priors were put on all but one parameter. Our simulations in Section~S1 of the Supplementary Material demonstrated an inherent difficulty of carrying out Bayesian inference for this (fine-scale) variance parameter, so we substituted its EM estimate into the hierarchical model. We carried out a sensitivity study on inference from the proposed hierarchical statistical model under different plug-in values of $\sigma^2_{\xi}$, and we found that the inference was not very sensitive to its misspecification. Our simulations in Section~S1 also allowed us to do some comparisons between the EHM and our (hybrid) BHM, and we found that BHM is preferred over EHM when making predictions in gaps of the spatial domain where no nearby observations are available. 

It is worth emphasizing that the sea-ice-extent data are obtained from the sea-ice concentrations, which are observed areal proportions that are noisy (especially for lower concentration percentages). Here, we have proposed a Bayesian spatio-temporal model for the binary SIE data, from which the predictive distribution of the underlying process on the probability scale is less uncertain than the observed proportions. The statistical dependencies due to spatial proximity and through time are modeled explicitly, and Bayesian inferences account for them naturally and coherently. Simple summaries based on an empirical approach cannot capture the spatio-temporal variations simultaneously with proper uncertainty quantification, due to statistical dependencies in space and time.

A new aspect of our article is the choice of various functionals of the latent GP. These summary statistics feature the process dynamics and, through their predictive distributions, they illustrate the changing patterns of Arctic sea ice between the earlier decade and the recent decade. For example, we consider boxplots of the posterior means of the latent GP on the original scale (and on the probability scale), empirical temporal semivariograms, Arctic-region H\"{o}vmoller diagrams, and spatial maps of loss and gain of ice, which offer different ``views" of the changing spatio-temporal patterns of Arctic sea ice. Two functionals that lend themselves to animation are the ice-to-water transition probabilities and the water-to-ice transition probabilities \citep{zhang2018estimating}. Still shots of that animation are given in Section~S5 of the Supplementary Material. 

\begin{figure}[ht!]
\centering
\includegraphics[width=12cm, height=8cm]{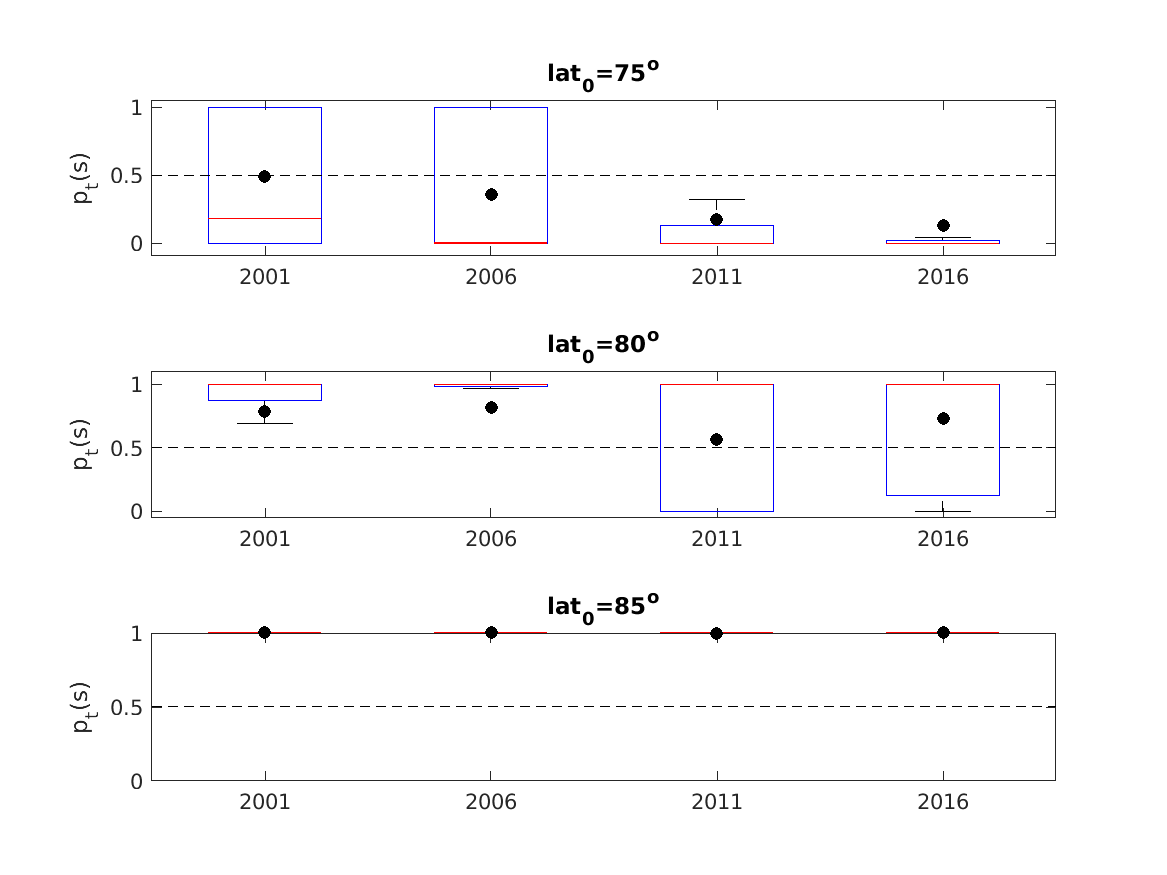}
\caption{Boxplots of $\{E(p_t(\mb{s})|\textnormal{data}): \mb{s}=(\textnormal{lon, lat})\ \textnormal{and}~\textnormal{lat}\in (\textnormal{lat}_0-\Delta, \textnormal{lat}_0+\Delta)\}$ for three latitude bands and for years $t=2001, 2006, 2011,2016$, where the width of the latitude band is $2\Delta=1^{\circ}$. The dashed line indicates the value of 0.5. }\label{Fig_bp_pscale}
\end{figure}

Figure~\ref{Fig_bp_pscale} shows boxplots of the posterior means of the latent probability of ice, denoted as $\{p_t(\mb{s})\}$, where the boxplots are taken at times $t=2001$, $2006$, $2011$, and $2016$, for $\mb{s}$ ranging over the latitude bands centered at $\textnormal{lat}_0=75^{\circ}$N, $80^{\circ}$N, and $85^{\circ}$N. The effects of climate change in the recent decade is seen in different ways, at different latitudes: For $\textnormal{lat}_0=75^{\circ}$N, the boxplot averages (denoted by a dot) decrease over time, dropping well below $0.5$ (the cut-off where there are more water pixels than ice pixels) during the recent decade. The high variability in the earlier decade, as captured by the size of the boxes, indicates a process in transition before collapse of the sea ice in the recent decade at that latitude. At $80^{\circ}$N, there is also a difference of the average posterior-mean $p_t(\mb{s})$-values between the earlier decade and the recent decade, and noticeably the variabilities are strikingly different. The high variability in the recent decade may precede a collapse as it did at $75^{\circ}$N. For the very high latitude of $85^{\circ}$N, all the average posterior-mean $p_t(\mb{s})$-values are very close to $1$ with very small variability, due to still-dominant ice cover at very high latitudes. Other useful functionals based on the predictive distribution of $\{p_t(\mb{s})\}$ are given in Section~\ref{Sec_RD_Analysis}, including temporal semivariograms that are able to detect a periodicity in the latent process at $85^{\circ}$N. 

The rest of the article is organized as follows. In Section~\ref{Sec_model}, we introduce a spatio-temporal hierarchical model for binary data, which is used for analyzing the Arctic sea-ice-extent (SIE) data. This consists of a data model, a process model that is dynamic, and a parameter model (or prior). An important part of the process model that we build is the large-scale variation captured by covariates in the process model; Section~\ref{Sec_model} also contains a detailed description of how the covariates were selected. In Section~\ref{Sec_Bayes}, we discuss Bayesian inference of model parameters and the prediction/forecasting procedure of the BHM. A Bayesian analysis of $20$ years of binary Arctic SIE data is given in Section~\ref{Sec_RD_Analysis}, where several summary statistics and their predictive distributions are used to detect the changing patterns of Arctic sea ice from 1997 \textendash~2016. In Section~5, a summary and discussion of our results are given. The article also contains five sections of Supplementary Material, to support our work presented in the main paper. 

\section{Spatio-Temporal Hierarchical Modeling}  \label{Sec_model}
In this section, we introduce a spatio-temporal hierarchical statistical model for the binary Arctic SIE data introduced in Section~1. Let $z_t(\mb{s})\in \{0, 1\}$ be a binary datum observed at a spatial location $\mb{s}\in\mathcal{D}_t$ and a time point $t\in \mathcal{T}$, where $\mathcal{D}_t$ is the spatial domain of interest and $t$ is a discrete time point in $\mathcal{T}\equiv\{1,2,\ldots, T\}$. We assume there is a latent process $\{y_t(\mb{s})\}$ that is used to define the probability that $z_t(\mb{s})$ equals $0$ or $1$, as follows. Following \cite{diggle1998model}, we model $\{z_t(\mb{s})\}$ as conditionally independent Bernoulli random variables given the latent process $\{y_t(\mb{s})\}$. That is, we model
\begin{eqnarray}
z_t(\mb{s})|y_t(\mb{s})\sim \mathrm{Ber}(p_t(\mb{s})), \label{Model_Response}
\end{eqnarray}
independently for all $\mb{s}\in \mathcal{D}_t$ and $t\in \mathcal{T}$, where $\mathrm{Ber}(p)$ is a binary random variable on $\{0, 1\}$ and equals $1$ with probability $p$.  In \eqref{Model_Response}, $y_t(\mb{s})=g(p_t(\mb{s}))$ and $g(\cdot)$ is a link function; here we choose the logit link such that $y_t(\mb{s})=g(p_t(\mb{s}))=\log(p_t(\mb{s})/(1-p_t(\mb{s})))$. Equation~\eqref{Model_Response} above defines the so-called ``data model.” 

The ``process model" is underneath the data model in the hierarchy of conditional models. Here, the model for $\{y_t(\mb{s})\}$, conditional on the parameters, is a spatio-temporal linear mixed model with a vector autoregressive model of order 1 (VAR(1)) for the coefficients of a relatively small set of basis functions \citep[e.g.,][]{wikle2001spatiotemporal,Cressie:Shi:Kang:2010,katzfuss2011spatio,bradley2015multivariate}: 
\begin{eqnarray}
&&y_t(\mb{s})=\mb{x}_t(\mb{s})^{\prime}\sbf{\beta}+\mb{S}_t(\mb{s})^{\prime}\sbf{\eta}_t+\xi_t(\mb{s}), \nonumber\\
&& \sbf{\eta}_t=\mb{H}_t\sbf{\eta}_{t-1}+\sbf{\zeta}_t, \label{Model_Process1}\\
&& \sbf{\zeta}_t\sim \textnormal{Gau}(\mb{0}, \mb{U}_t)\  \textnormal{and} \  
\sbf{\eta}_1\sim \textnormal{Gau}(\mb{0}, \mb{K}),\nonumber
\end{eqnarray}
where at time $t\in \mathcal{T}$ and $\mb{s}\in \mathcal{D}_t$, $\mb{x}_t(\mb{s})$ is a $p$-dimensional vector of covariates at spatial location $\mb{s}$; $\sbf{\beta}$ is the $p$-dimensional vector of regression coefficients associated with $\mb{x}_t(\mb{s})$; $\mb{S}_t(\mb{s})$ is an $r$-dimensional basis-function vector evaluated at $\mb{s}$; $\sbf{\eta}_t$ is a vector of random coefficients of $\mb{S}_t(\mb{s})$; $\xi_t(\mb{s})$ is a random variable that models the fine-scale variation at $\mb{s}$ not captured by $\mb{S}_t(\mb{s})^{\prime}\sbf{\eta}_t$; and $\textnormal{Gau}(\mu, \sigma^2)$ denotes a Gaussian distribution with mean $\mu$ and variance $\sigma^2$. The parameters in \eqref{Model_Process1}, upon which the process model is conditioned, are: $\sbf{\beta}$, the $p$-dimensional vector of regression coefficients; and $\{\mb{H}_t: t=2,\ldots, T\}$ and $\{\mb{U}_t: t=2,\ldots, T\}$, the $r\times r$ propagator and $r\times r$ innovation matrices, respectively. We further assume that $\xi_t(\mb{s})$ in \eqref{Model_Process1} follows $\textnormal{Gau}(0, \sigma^2_{\xi})$ distribution, $\{\xi_t(\mb{s})\}$ are independent of each other over both space and time, and they are also independent of the random vectors $\{\sbf{\eta}_t: t=1, \ldots, T\}$. Hence, spatio-temporal variability in the data is captured through the fixed-effects term, $\mb{x}_t(\mb{s})^{\prime}\sbf{\beta}$ (large-scale variation), and the random-effects term, $\mb{S}_t(\mb{s})^{\prime}\sbf{\eta}_t$ (small-scale variation), and \eqref{Model_Process1} represents a spatio-temporal GP. 

For a fixed time point $t$, the number of observations can be very large; that is, $\{y_t(\mb{s}): \mb{s}\in \mathcal{D}_t\}$ forms a high-dimensional vector when evaluated at all the observation locations. By fixing the number of basis functions $r$ to be a relatively small number (say a few hundred), the latent process, $\{y_t(\mb{s})-\mb{x}_t(\mb{s})^{\prime}\sbf{\beta}\}$, is represented by a low-dimensional basis-function vector, allowing fast computations to be achieved \citep[e.g.,][]{wikle2001spatiotemporal,cressie2006spatial,cressie2008frk,kang2010using,zammit2019frk}. There are many types of basis functions that could be used in this setting, such as wavelets, splines, Wendland functions, and bisquare functions. In this article, we focus on the compactly supported bisquare functions, since they have been successfully used to model very large Gaussian and non-Gaussian spatial and spatio-temporal data \citep[e.g.,][]{cressie2008frk,sengupta2013hierarchical,zhang2018estimating}. In addition, specifying multi-resolution basis functions has proven to be effective in capturing spatial dependence at different scales \citep[e.g., ][]{wikle2001spatiotemporal,katzfuss2011spatio,nychka2015multiresolution,katzfuss2017multi}. Hence, we adopt a multi-resolution class.

For $j=1,\ldots, r_i$ basis functions of the $i$-th resolution, we define the bisquare basis function in $d$-dimensional Euclidean space $\mathbb{R}^d$ as,
\begin{eqnarray}
S_j^{(i)}(\mb{s})\equiv\left(1-\left(\frac{\|\mb{s}-\mb{c}_j^{(i)}\|}{\phi_i}\right)^2\right)^2I(\|\mb{s}-\mb{c}_j^{(i)}\|<\phi_i);~\mb{s}\in\mathbb{{R}}^d, \label{BF_bisquare}
\end{eqnarray}
where $r_i$ is the number of basis functions at the $i$-th resolution, $\mb{c}_j^{(i)}$ is the center of the $j$-th basis function $S_j^{(i)}(\cdot)$ at the $i$-th resolution, $\|\cdot\|$ is the Euclidean norm, $\phi_i$ is the radius of its spatial support (sometimes called the aperture), and $I(\cdot)$ is an indicator function. In practice, $\phi_i$ is specified to be $1.5$ times the minimum distance between basis-function centers of the same resolution \citep[e.g.,][]{cressie2008frk}. Note that in Section~\ref{Sec_RD_Analysis}, we replace the Euclidean norm with the great-circle distance in \eqref{BF_bisquare}, and it is easy to see that this type of modification could be made on any manifold equipped with a norm.

The propagator matrix $\mb{H}_t$ captures the temporal covariances between the elements of $\sbf{\eta}_t$ and $\sbf{\eta}_{t-1}$, for $t=2, \ldots, T$. Here we treat $\mb{H}_t$ and $\mb{U}_t$ as unknown parameters, but we assume that they are constant for a fixed time period; that is, $\mb{H}_t\equiv \mb{H}$ and $\mb{U}_t\equiv \mb{U}$, for $t=2,\ldots, T$. In practice, they will be allowed to vary from time-period to time-period \citep[e.g.,][]{katzfuss2011spatio,zhang2018estimating}. When modeling the Arctic SIE data, we allow the two matrices to change in successive five-year periods over the two decades of data. 

Although the propagator matrix $\mb{H}_t$ may be considered as an $r\times r$ parameter matrix, a parsimonious representation for it can improve inference and allow for physical interpretations of its entries \citep{wikle2001spatiotemporal}. Alternatively, \cite{bradley2015multivariate} proposed a class they called the Moran’s I class of propagator matrices in order to avoid confounding between $\{\sbf{\eta}_t\}$ and the covariates. Here we specify $\mb{H}_t$ for the multi-resolution basis functions by assigning a different propagator parameter for each resolution. For example, for a two-resolution design with $r_1$ Resolution-1 (coarse resolution) basis functions and $r_2$ Resolution-2 (fine resolution) basis functions, \cite{zhang2018estimating} parameterized $\mb{H}_t$ as follows: For $t=2,\ldots, T$,
\begin{eqnarray}
\mb{H}_t\equiv \mb{H}\equiv\left(\begin{array}{cc}\lambda_{1} \mb{I}_{r_1}& \mb{0}\\ \lambda_{3}\mb{R}& \lambda_{2} \mb{I}_{r_2}\end{array}\right), \label{Parameterization_H}
\end{eqnarray}
where $\lambda_1,\lambda_2,\lambda_3\in (-1, 1)$, and $\mb{R}$ is an $r_2\times r_1$ matrix encoding the possible dependence from the coarse Resolution 1 to the fine Resolution 2. Specifically, for $i=1,\ldots, r_2$ and $j=1,\ldots, r_1$, $\mb{R}(i,j)=1$ if the $i$-th Resolution-2 basis function is a neighbor of the $j$-th Resoluton-1 basis function, and $\mb{R}(i,j)=0$ otherwise. If the model is used for forecasting, we would need to check the eigenvalues of $\mb{H}$ for possible ``explosive" behavior. 

The parameters for the data model given by \eqref{Model_Response} and the parameters for the process model given by \eqref{Model_Process1} are $\sbf{\theta}\equiv\{\sbf{\beta},\sigma^2_{\xi},\mb{K},\mb{H},\mb{U}\}$. For Bayesian inference, we need to assign a prior to $\sbf{\theta}$: For variance-covariance parameters $\mb{K}$, $\mb{U}$, and $\sigma^2_{\xi}$, we used Inverse-Wishart distributions for conditional conjugacy; for the regression coefficients $\sbf{\beta}$, we specified an improper non-informative prior $\pi(\sbf{\beta})\propto 1$; and for each of the $\{\lambda_i\}$ of $\mb{H}$ given by \eqref{Parameterization_H}, we specified independent $\mathrm{Unif}(-1, 1)$ distributions. In the simulation setting given in Section~S1 of the Supplementary Material, we found that a single parameter, $\sigma^2_{\xi}$, that is better handled through EM estimation than through assigning it a prior, and hence we do likewise for analyzing the Arctic SIE data. 

\subsection{Specification of covariates for the Arctic sea-ice-extent data} \label{Sec_describe_covariates}

In this subsection, we discuss how to select covariates for modeling the spatio-temporal Arctic SIE data in the month of September for each year spanning the two decades from 1997~\textendash~2016. September is chosen as a time when Arctic sea ice is least, and it avoids modeling the within-year seasonal variation \citep{parkinson1999arctic}; consequently, we are not able to detect changes in seasonal variation over the years. Henceforth $\mathcal{D}_t\equiv \mathcal{D}$, a region defined by pixels whose latitudes are greater than or equal to $60^{\circ}$N that we call the Arctic region. 

\begin{figure}[!ht]
\centering
\includegraphics[width=12cm, height=6cm]{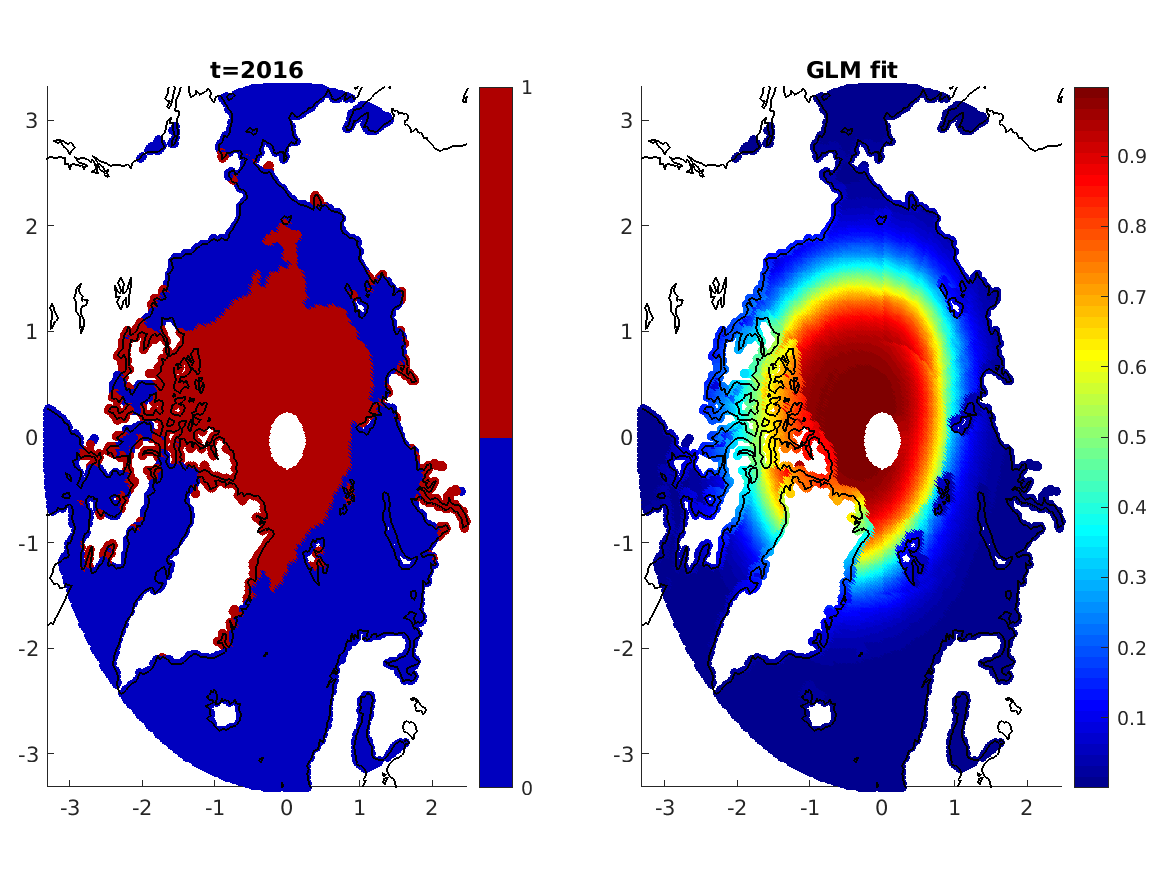}
\caption{September Arctic SIE data at $t=2016$ (binary, left panel) and the GLM-fitted logistic regression fitted to the 2016 data (using the covariates in \eqref{Eqn_mean_SIE}) on the probability scale (right panel). The white circle around the North Pole in both panels represents a region not covered by the sensor due to remote sensing limitations.} \label{Fig_EDA_1997_mean}
\end{figure}

Recall that in \cite{zhang2018estimating}, the only covariate used was the distance of any spatial location in $\mathcal{D}$ to the North Pole, which is a proxy for surface temperatures in the Arctic region; its coefficient was allowed to vary with $t\in \mathcal{T}$ in order to detect climate change. In this article, we consider the Arctic surface temperatures directly and use the GISTEMP surface-temperature-anomaly data to obtain covariates. Specifically, we use the previous year's summer-average and winter-average surface-temperature anomalies to define two covariates that allow for temporal variation as well as spatial variation. In addition, for some spatial pixels that are close to coastlines but far from the North Pole, we observe the presence of sea ice (e.g., see the left panel of Figure~\ref{Fig_EDA_1997_mean}). Hence, we use the distance of spatial location to the nearest coastline to define a covariate that models local spatial effects. Finally, an exploratory data analysis based on a simple logistic-regression model showed that longitude played a role in modeling these data, probably due to the different dynamics in different regions of the Arctic (Section~S2 of the Supplementary Material).

Based on preliminary data analyses that included use of standard logistic regressions, we fitted the following mean function for the hidden spatio-temporal GP:
\begin{eqnarray}
\mb{x}_t(\mb{s})^{\prime}\sbf{\beta}&=&\beta_0+\bar{x}^{su}_{t-1}\beta_1+\bar{x}^{wi}_{t-1}\beta_2+(x^{su}_{t-1}(\mb{s})-\bar{x}^{su}_{t-1})\beta_3+(x^{wi}_{t-1}(\mb{s})-\bar{x}^{wi}_{t-1})\beta_4 \nonumber \\
&&\qquad+\cos(\pi s_1/180)\beta_5+\sin(\pi s_1/180)\beta_6+x^{pl}(\mb{s})\beta_7+x^{cs}(\mb{s})\beta_8,  \label{Eqn_mean_SIE}
\end{eqnarray}
where $\mb{s}\equiv (s_1, s_2)^{\prime}\in \mathcal{D}$, and $s_1$ and $s_2$ are the longitude and latitude of $\mb{s}$; $\bar{x}^{su}_{t-1}$ and $\bar{x}^{wi}_{t-1}$ are the spatially averaged (i.e., over $\mathcal{D}$) Boreal summer (averaged over Jun, Jul, Aug) and winter (averaged over Dec, Jan, Feb) surface-temperature anomalies each indexed by the year $(t-1)$ in which the first month falls. The components of the spatial averages $x^{su}_{t-1}(\mb{s})$ and $x^{wi}_{t-1}(\mb{s})$ at the spatial location $\mb{s}$ and the previous year $(t-1)$ are used to capture the regional effects of changes of surface temperatures. That is, summer (winter) in year $(t-1)$ corresponds to the three-month season starting in Jun (Dec) in year $(t-1)$. These first four covariates aim to characterize the effects of the previous year's surface-temperature anomalies on forming the sea ice in September in the current year. 

The remaining four covariates are purely spatial: $x^{pl}(\mb{s})$ is the distance between $\mb{s}$ and the North Pole and was the only covariate included in the hidden spatio-temporal GP of the EHM fitted by \cite{zhang2018estimating}; the covariates based on longitude represent periodic spatial heterogeneity, analogous to modeling a linear trend on the real line; and $x^{cs}(\mb{s})$ is a more local spatial effect accounting for short distances to the coast (less than $50$km). Let $d^{cs}(\mb{s})$ denote the great-circle distance from $\mb{s}$ to the nearest coastline. Our exploratory data analysis based on logistic regression showed that the coefficients of $x^{pl}(\mb{s})$ are larger for data close to the coastline than for data far away from the coastline. Then we define $x^{cs}(\mb{s})\equiv x^{pl}(\mb{s})\times I(d^{cs}(\mb{s})<50 \textnormal{km})$ to provide adjustments for the effect of distance to the coast. The right panel of Figure~\ref{Fig_EDA_1997_mean} shows the logistic-regression fit of \eqref{Eqn_mean_SIE} for the Arctic sea-ice-extent data for $t=2016$. 

In summary, we shall use the eight covariates in \eqref{Eqn_mean_SIE} to model the GP's fixed effects in Section~\ref{Sec_RD_Analysis}, and the remaining spatio-temporal variability in the GP is modeled statistically with random effects given in \eqref{Model_Process1}. 

\section{Fully Bayesian Inference}\label{Sec_Bayes}

In this section, we discuss the Bayesian inference of model parameters for the hierarchical statistical model given by \eqref{Model_Response} and \eqref{Model_Process1}. Let $\mathcal{D}_t^o\equiv\{\mb{s}_{t,1}^o, \mb{s}_{t,2}^o, \ldots, \mb{s}_{t, N_t}^o\}$ be the spatial locations with observed data at time $t\in\{1,\ldots, T\}$. For each time $t$, we stack the data into a column vector, $\mb{Z}_t^o\equiv(z_t(\mb{s}_{t,1}^o), \ldots, z_t(\mb{s}_{t,N_t}^o))^{\prime}$, and we define $\sbf{\xi}_t^o\equiv(\xi_t(\mb{s}_{t,1}^o), \ldots, \xi_t(\mb{s}_{t,N_t}^o))^{\prime}$ to be the fine-scale-variation vector evaluated at $\mathcal{D}_t^o$. Let $\mb{Z}^{o}\equiv(\mb{Z}_1^{o^{\prime}}, \ldots, \mb{Z}_T^{o^{\prime}})^{\prime}$, $\sbf{\xi}^o\equiv(\sbf{\xi}_1^{o^{\prime}}, \ldots, \sbf{\xi}_T^{o^{\prime}})^{\prime}$, and $\sbf{\eta}\equiv(\sbf{\eta}_1^{\prime}, \ldots, \sbf{\eta}_T^{\prime})^{\prime}$; then the data likelihood is:
\begin{eqnarray*}
L(\sbf{\theta};\mb{Z}^o)&=&\int_{\sbf{\eta}}\int_{\sbf{\xi}^o} p(\mb{Z}^o|\sbf{\eta}, \sbf{\xi}^o, \sbf{\beta})\times p(\sbf{\eta}|\mb{K}, \mb{H}, \mb{U})\times p(\sbf{\xi}^o|\sigma^2_{\xi})d\sbf{\xi}^o d\sbf{\eta}\nonumber\\
&=&\int_{\sbf{\eta}}\int_{\sbf{\xi}^o}\prod\limits_{t=1}^T\prod\limits_{i=1}^{N_t} (1+\exp(-(2z_{t,i}^o-1)y_{t,i}^o))^{-1}\times (2\pi)^{-r/2}|\mb{K}|^{-1/2}\exp(-\sbf{\eta_1}^{\prime}\mb{K}^{-1}\sbf{\eta}_1/2)\nonumber\\
&&\times (2\pi)^{-(T-1)r/2}|\mb{U}|^{-(T-1)/2}\prod\limits_{t=2}^T\exp(-(\sbf{\eta}_t-\mb{H}\sbf{\eta}_{t-1})^{\prime}\mb{U}^{-1}(\sbf{\eta}_t-\mb{H}\sbf{\eta}_{t-1})/2)\nonumber\\
&&\times \prod\limits_{t=1}^T(2\pi\sigma^2_{\xi})^{-N_t/2}\exp(-\sbf{\xi}_t^{o^{\prime}}\sbf{\xi}_t^o/(2\sigma^2_{\xi})) \mathrm{d}\sbf{\xi}^o \mathrm{d}\sbf{\eta}, 
\end{eqnarray*}
where $p(\cdot|\cdot)$ represents the appropriate conditional-probability density; under this notation, $p(\mb{Z}^o|\sbf{\eta}, \sbf{\xi}^o, \sbf{\beta})$ is a product of Bernoulli densities, $p(\sbf{\eta}|\mb{K}, \mb{H}, \mb{U})$ is a multivariate Gaussian density, and $p(\sbf{\xi}^o|\sigma^2_{\xi})$ is a product of univariate Gaussian densities. For notational simplicity, we define $z_{t,i}^o\equiv z_t(\mb{s}_{t,i}^o)$, and $y_{t,i}^o\equiv y_t(\mb{s}_{t,i}^o)$. Since the integration above 
does not have a closed form due to the nonlinearity of the data model, \cite{zhang2018estimating} modified the spatial-only methodology given in \cite{sengupta2013hierarchical} and developed an EM algorithm to estimate model parameters, $\sbf{\theta}$, for spatio-temporal data $\mb{Z}^o$. Then they substituted the resulting estimate $\hat{\sbf{\theta}}$ into the hierarchical model, resulting in an EHM. In this article, our approach is Bayesian, and (with the exception of one parameter, $\sigma^2_{\xi}$) we put prior distributions on the elements of $\sbf{\theta}$, resulting in a BHM and a joint posterior distribution of $\sbf{\theta}$, $\sbf{\eta}$, and $\sbf{\xi}^o$.  

In real-world applications, some of the parameters may be fixed by the modeler at physically meaningful values or at estimates, and the remaining parameters could have priors put on them. This results in what might be called a hybrid hierarchical model that is not an EHM but not quite a BHM either \citep[e.g.,][Ch.1]{Wikle_Mangion_Cressie_2019}. When analyzing the Arctic SIE data, $\sigma^2_{\xi}$ was very difficult to infer from a fully Bayesian analysis. This may be because it is hard to identify separately the small-scale variation and the fine-scale variation from the total variation observed on binary data, particularly when $\mb{K}$ is an arbitrary $r\times r$ positive-definite matrix. However, we found the situation unimproved by making parametric assumptions about $\mb{K}$ and, more fundamentally, this may be because not all the covariance parameters are identifiable under fixed-domain asymptotics \citep{zhang2004inconsistent}. 

In Section~4, we plug in the EM estimate of $\sigma^2_{\xi}$, resulting in a hybrid BHM, although we continue to call the model a BHM. However, in the rest of Section~\ref{Sec_Bayes}, we present a fully Bayesian methodology that includes a prior on $\sigma^2_{\xi}$, noting that the algorithms simplify a little when we fix $\sigma^2_{\xi}$. 

\subsection{Parameter inference} \label{Sec_Estimation}
Let $\pi(\cdot)$ denote a generic prior distribution. The joint posterior distribution, $p(\sbf{\eta}, \sbf{\xi}^{o}, \sbf{\theta}|\mb{Z}^o)$, is proportional to the product of the complete likelihood, $p(\mb{Z}^o, \sbf{\eta}, \sbf{\xi}^{o}| \sbf{\theta})$, and the prior distribution, $\pi(\mb{K},\mb{H}, \mb{U}, \sbf{\beta}, \sigma^2_{\xi})$.
For $\mb{H}$ parameterized as in \eqref{Parameterization_H} and $\textnormal{IW}(\nu, \sbf{\Phi})$ denoting an Inverse Wishart distribution with parameters $\nu$ and $\sbf{\Phi}$, we put $\pi(\mb{K})\sim \textnormal{IW}(\nu_K, \sbf{\Phi}_K)$, $\pi(\mb{U})\sim \textnormal{IW}(\nu_U, \sbf{\Phi}_U)$, $\pi(\lambda_i)\sim \textnormal{Unif}(-1, 1)$ independently for $i=1,2,3$, and $\pi(\sbf{\beta}, \sigma^2_{\xi})\propto (\sigma^2_{\xi})^{-1/2}$. Then we can integrate out $\mb{K}$ and $\mb{U}$ and sample $\{\sbf{\eta}, \sbf{\xi}^o, \sbf{\beta}, \{\lambda_i\}, \sigma^2_{\xi}\}$ from 
\begin{eqnarray}
\hspace{-0.35cm}p(\sbf{\eta}, \sbf{\xi}^{o}, \sbf{\beta}, \{\lambda_i\}, \sigma^2_{\xi}|\mb{Z}^o)&\propto& p(\mb{Z}^{o}|\sbf{\beta}, \sbf{\eta}, \sbf{\xi}^o)\times |\sbf{\eta}_1\sbf{\eta}_1^{\prime}+\sbf{\Phi_K}|^{-\frac{\nu_K+1}{2}}\times (\sigma^2_{\xi})^{-(N+1)/2}\times e^{-\frac{\sbf{\xi}^{o\prime}\sbf{\xi}^o}{2\sigma^2_{\xi}}}\nonumber\\
&&\times |\sum\limits_{t=2}^T(\sbf{\eta}_t-\mb{H}\sbf{\eta}_{t-1})(\sbf{\eta}_t-\mb{H}\sbf{\eta}_{t-1})^{\prime}+\sbf{\Phi_U}|^{-\frac{\nu_U+T-1}{2}}, \label{Eqn_post_integrated_out}
\end{eqnarray}
where $N=\sum_{t=1}^TN_t$. Note that we do not integrate out $\sigma^2_{\xi}$, because we would like to sample each $\xi_{t,i}^o$ individually such that the computations are parallelizable (and we note that ultimately we fix $\sigma^2_{\xi}$ at its EM estimate in Section~\ref{Sec_RD_Analysis}). The posterior samples of $\mb{K}$ and $\mb{U}$ can be sampled from the following full conditionals:
\begin{eqnarray}
&&p(\mb{K}|\cdot)\sim \textnormal{IW}(\nu_K+1, \sbf{\eta}_1\sbf{\eta}_1^{\prime}+\sbf{\Phi_K}), \label{Eqn_post_K}\\
&&p(\mb{U}|\cdot)\sim \textnormal{IW}\left(\nu_U+T-1, \sum\limits_{t=2}^T(\sbf{\eta}_t-\mb{H}\sbf{\eta}_{t-1})(\sbf{\eta}_t-\mb{H}\sbf{\eta}_{t-1})^{\prime}+\sbf{\Phi_U}\right), \label{Eqn_post_U}
\end{eqnarray}
where ``$\cdot$" denotes conditioning on all other parameters (and of course on the data). In order to make $\{\sbf{\eta}_t: t=1,\ldots, T\}$ well constrained in the prior when modeling the Arctic SIE data, we chose the Inverse Wishart parameter $\nu_K=2r$. This makes the prior distribution more informative; then $\sbf{\Phi}_{K}=(3r+1)\hat{\mb{K}}$ and $\sbf{\Phi}_{U}=(3r+1)\hat{\mb{U}}$, to make the priors of $\mb{K}$ and $\mb{U}$ concentrate around their respective EM estimates, denoted as $\hat{\mb{K}}$ and $\hat{\mb{U}}$.

For $\sbf{\beta}$, $\{\lambda_i\}$, $\sbf{\eta}$, and $\sbf{\xi}^o$, whose full conditionals do not have a closed form, we used a Metropolis-Hastings algorithm within the Gibbs sampler \citep{gelfand1990sampling,gelman2014bayesian} to draw their posterior samples. Following \cite{zhang2018estimating}, we generated the posterior samples of $\{\sbf{\eta}_t\}$ successively; for $\sbf{\xi}^o$, we generated the posterior samples of $\{\xi_{t, i}^o\}$ individually. Details of the MCMC sampling algorithm and convergence diagnostics for the Arctic SIE data are given in Section~S3 of the Supplementary Material. 

\subsection{Prediction and forecasting} \label{Section_prediction_forecasting}

Let $\mathcal{D}_t^u\equiv\{\mb{s}_{t,1}^u, \mb{s}_{t,2}^u, \ldots, \mb{s}_{t, m_t}^u\}$ be the spatial locations specified for prediction at time $t\in\{1,\ldots, T\}$, and let $\mb{Y}_{t}^u\equiv (y_t(\mb{s}_{t,1}^u), \ldots, y_t(\mb{s}_{t, m_t}^u))^{\prime}$ denote the vector at those prediction locations. Note that the prediction-location set $\mathcal{D}_t^u$ and the observation-location set $\mathcal{D}_t^o$ may have a non-empty intersection. If $\mathcal{D}_t^u$ and $\mathcal{D}_t^o$ overlap, which is our interest here when modeling the Arctic SIE data, then for location $\mb{s}^o_{t,i}\in \mathcal{D}_t^o$, we plug in the posterior samples of $\sbf{\eta}_t$, $\xi_t(\mb{s}_{t,i}^o)$, and $\sbf{\theta}$ into the process-model equation, $y_t(\mb{s}^o_{t,i})=\mb{x}_t(\mb{s}^o_{t,i})^{\prime}\sbf{\beta}+\mb{S}_t(\mb{s}^o_{t,i})^{\prime}\sbf{\eta}_t+\xi_t(\mb{s}^o_{t,i})$, to obtain the samples from the predictive distribution of $y_t(\mb{s}^o_{t,i})$, from which summaries (e.g., mean, variance, and quantiles) of $y_t(\mb{s}^o_{t,i})$ can be readily obtained. 

Now consider the scenario that the prediction-location set $\mathcal{D}_t^u$ and the observation-location set $\mathcal{D}_t^o$ do not overlap, which is the case when predicting in the region around the North Pole where no observations are available. Now samples from the predictive distribution can be obtained based on the posterior samples of $\sbf{\eta}$ and $\sbf{\theta}$. Since
\begin{eqnarray}
p(\mb{Y}_t^u|\mb{Z}^o)&=&\int p(\mb{Y}_t^u|\sbf{\eta}_t, \sbf{\theta}, \mb{Z}^o)\times p(\sbf{\eta}_t, \sbf{\theta}|\mb{Z}^o)\mathrm{d}\sbf{\eta}_t\mathrm{d}\sbf{\theta} \nonumber\\
&=&\int p(\mb{Y}_t^u|\sbf{\eta}_t, \sbf{\theta})\times p(\sbf{\eta}_t, \sbf{\theta}|\mb{Z}^o)\mathrm{d}\sbf{\eta}_t\mathrm{d}\sbf{\theta}, \label{Eqn_Pred_Y}
\end{eqnarray}
the predictive samples of $\mb{Y}_t^u$ can be drawn by the method of composition: Having drawn the posterior samples of $\sbf{\eta}_t$ and $\sbf{\theta}$, $\mb{Y}_t^u$ is drawn from $p(\mb{Y}_t^u|\sbf{\eta}_t, \sbf{\theta})$, which is a Gaussian distribution. The predictive mean and predictive variance of $\mb{Y}_t^u$ are then obtained from: 
\begin{eqnarray*}
E(\mb{Y}_t^u|\mb{Z}^o)&=&E(E(\mb{Y}_t^u|\sbf{\eta}_t, \sbf{\theta})|\mb{Z}^o)=\mb{X}_t^uE(\sbf{\beta}|\mb{Z}^o)+\mb{S}_t^uE(\sbf{\eta}_t|\mb{Z}^o), \\
\textnormal{var}(\mb{Y}_t^u|\mb{Z}^o)&=& E(\textnormal{var}(\mb{Y}_t^u|\sbf{\eta}_t, \sbf{\theta})|\mb{Z}^o)+\textnormal{var}(E(\mb{Y}_t^u|\sbf{\eta}_t, \sbf{\theta})|\mb{Z}^o) \\
&=&E(\sigma^2_{\xi}|\mb{Z}^o)\mb{I}_{m_t}+\textnormal{var}(\mb{X}_t^u\sbf{\beta}+\mb{S}_t^u\sbf{\eta}_t|\mb{Z}^o),
\end{eqnarray*}
where $\mb{I}_{m_t}$ is the $m_t\times m_t$ identity matrix, $\mb{X}_t^u\equiv (\mb{x}_t(\mb{s}_{t,1}^u), \ldots, \mb{x}_t(\mb{s}_{t,m_t}^u))^{\prime}$, and $\mb{S}_t^u\equiv (\mb{S}_t(\mb{s}_{t,1}^u), \ldots, \mb{S}_t(\mb{s}_{t,m_t}^u))^{\prime}$.

Given the predictive samples of $\mb{Y}^u$, then for $\mb{Z}^u_t\equiv (z_t(\mb{s}_{t,1}^u), \ldots, z_t(\mb{s}_{t, m_t}^u))^{\prime}$, its predictive samples are drawn from 
\begin{eqnarray}
p(\mb{Z}^u_t|\mb{Z}^o)=\int p(\mb{Z}^u_t|\mb{Y}^u_t, \mb{Z}^o)\times p(\mb{Y}_t^u|\mb{Z}^o) \mathrm{d}\mb{Y_t^u}=\int p(\mb{Z}^u_t|\mb{Y}^u_t)\times p(\mb{Y}_t^u|\mb{Z}^o) \mathrm{d}\mb{Y_t^u},\label{Eqn_Pred_Z}
\end{eqnarray}
again by the method of composition, using the posterior samples of $\mb{Y}_t^u$. The predictive mean and predictive variance of $\mb{Z}^u_t$ are 
\begin{eqnarray*}
E(\mb{Z}_t^u|\mb{Z}^o)&=&E(E(\mb{Z}_t^u|\mb{Y}_t^u)|\mb{Z}^o)=E(g^{-1}(\mb{Y}_t^u)|\mb{Z}^o),\\
\textnormal{var}(\mb{Z}_t^u|\mb{Z}^o)&=&E(\textnormal{var}(\mb{Z}_t^u|\mb{Y}_t^u)|\mb{Z}^o)+\textnormal{var}(E(\mb{Z}_t^u|
\mb{Y}_t^u)|\mb{Z}^o)\\
&=&E(\textnormal{diag}\{g^{-1}(\mb{Y}_t^u)(1-g^{-1}(\mb{Y}_t^u))\}|\mb{Z}^o)+\textnormal{var}(g^{-1}(\mb{Y}_t^u)|\mb{Z}^o),
\end{eqnarray*}
where $\textnormal{diag}\{\mb{a}\}$ denotes a diagonal matrix with its diagonal entries given by the vector $\mb{a}$, and recall that $g(\cdot)$ is the logit link function. 

When $t=T+1$, that is, when forecasting one-year ahead, we use the fact that $p(\sbf{\eta}_{T+1},\sbf{\theta}|\mb{Z}^o)=\int p(\sbf{\eta}_{T+1}|\sbf{\eta}_T, \sbf{\theta})\times p(\sbf{\eta}_T, \sbf{\theta}|\mb{Z}^o)\mathrm{d}\sbf{\eta}_T$, and equation \eqref{Eqn_Pred_Y}, to obtain the one-step-ahead forecast of $\mb{Y}_{T+1}^u$. Using equation \eqref{Eqn_Pred_Z}, $\mb{Z}_{T+1}^u$ can be similarly obtained.  

A simulation study comparing parameter estimation and prediction performance of the EHM and the BHM is given in Section~S1 of the Supplementary Material. From the simulation results, we found that when there are a limited number of observations, the EM estimates of the regression coefficients can have a large bias, but these parameters were estimated reasonably well by the BHM. We also found that the BHM yielded a considerably smaller prediction error than that of the EHM when predicting in spatial gaps where no nearby observations are available. Also in Section S1, a sensitivity study on the inference of the proposed BHM was conducted for different plug-in values of $\sigma^2_{\xi}$. We observed that the inference results were very similar and, for the different plug-in values, the $95\%$ credible intervals of the model parameters covered their respective true values. 

\section{Bayesian Analysis of the Arctic Sea-Ice-Extent Data} \label{Sec_RD_Analysis}

In this section, we fitted the spatio-temporal hierarchical statistical model developed in Section~\ref{Sec_model} and Section~\ref{Sec_Estimation} to Arctic sea-ice-extent data obtained from remote sensing from 1997 to 2016 (20 years in total). The temporal domain is $t=1997, \ldots, 2016$, which indexes the sea-ice data in the month of September for each of the 20 years. The original data are on a $304\times 448$ longitude-latitude grid, involving $136,192$ monthly observations. Since the grid cells (pixels) with latitude smaller than $60^{\circ}$N are always water pixels in the month of September, we focus on the spatial domain that covers the observation locations with latitude $60^{\circ}$ and above, resulting in the Arctic region from the south end of Greenland to the North Pole; see the left panel of Figure~\ref{Fig_EDA_1997_mean}. In total, there are $26, 342$ observation locations, and these spatial locations stay the same over time (i.e., $\mathcal{D}_t\equiv \mathcal{D}$). 

We partitioned the entire 20 years into four time periods: 1997 \textendash~2001, 2002 \textendash~2006, 2007 \textendash~2011, and 2012 \textendash~2016, in order to allow the dimension-reduced propagator matrix $\mb{H}$ and the corresponding innovation matrix $\mb{U}$ to vary over different time periods. That is, we fitted the BHM \eqref{Model_Response} \textendash~\eqref{Eqn_mean_SIE} with priors given in Section~\ref{Sec_Estimation}, but $\sigma^2_{\xi}$ was estimated (EM estimation) from the data in each time period separately and substituted into the BHM. Note that the end year of the previous time period was used to initialize the analysis of the current time period: Specifically, $\mb{K}_t=\mb{H}\mb{K}_{t-1}\mb{H}^{\prime}+\sigma^2_{\xi}\mb{I}_{N_t}$, for $t=2,\ldots, T$, where $\mb{K}_1\equiv \mb{K}, \mb{H}$, and $\sigma^2_{\xi}$ were obtained from the previous time period. This helps to avoid artificial abrupt transitions when inferring $\{y_t(\mb{s})\}$.

The surface-temperature-anomaly data that were used to define covariates $\mb{x}_t(\mb{s})$ in \eqref{Eqn_mean_SIE} come from the Goddard Institute for Space Studies Surface Temperature Analysis (GISTEMP) project \citep[e.g.,][]{Hansen2010Global,data_gisstemp}. This data set is on a $2^{\circ}\times 2^{\circ}$ longitude-latitude grid, and hence its resolution is coarser than the spatial grid of the Arctic SIE data, which has a resolution of approximately $25\textnormal{km}\times25\textnormal{km}$. To each sea-ice-observation location (i.e., center of the grid cell), we assigned its surface-air-temperature-anomaly value to be the closest surface-air-temperature-anomaly datum in the GISTEMP data set. Recall that we use eight covariates in \eqref{Eqn_mean_SIE} for modeling the fixed-effects term for the latent process $\{y_t(\mb{s})\}$, all of which are defined at all observation locations by this interpolation method. Further discussion of the covariates is given in the next subsection. 

\subsection{BHM fitting}

Following \cite{zhang2018estimating}, we used a two-resolution design for the basis functions. We defined $r_1 = 45$ Resolution-1 basis-function centers and $r_2 = 172$ Resolution-2 basis-function centers on regularly spaced geodesic grids on the polar cap. Some basis-function centers were placed outside the study domain to account for boundary effects \citep{cressie2010high}. Specifically, the Matlab function {\sf GridSphere} \citep{matlab_grid_sphere} was used to generate basis-function centers of two resolutions, where the great-circle radii of the Resolution-1 and Resolution-2 basis functions are $881.71\textnormal{km}$ and $440.86\textnormal{km}$, respectively. The bisquare basis functions in \eqref{BF_bisquare} were used to form $\{\mb{S}_t(\mb{s})^{\prime}\sbf{\eta}_t\}$, the small-scale-variation component of $\{y_t(\mb{s})\}$, and the basis-function matrices are $\mb{S}_t\equiv \mb{S}$ for each time period. We then fitted the proposed dimension-reduced Bayesian ST-GLMM to the Arctic SIE data for Periods 1~\textendash~4, with the fine-scale variance $\sigma^2_{\xi}$ in each given period fixed at its EM estimate. 

{\begin{table}[!ht]
\caption{Posterior means and $95\%$ credible intervals (in parentheses) for scalar parameters from fitting a BHM to the Arctic SIE data. The fine-scale variance $\sigma^2_{\xi}$ is fixed at its EM estimate in each time period.}
\label{Table_RD_PE}
\centering
\vspace{8pt}
\scalebox{0.65}{\begin{tabular}{|c|c|c|c|c|}
\toprule
Parameter & Period 1 & Period 2 & Period 3& Period 4\\
\hline
$\beta_0$& $-1.298~(-1.392, -1.205)$& $-2.284~(-2.339, -2.226)$& $-3.416~(-3.483, -3.349)$&$-2.748~(-2.811, -2.686)$\\
$\beta_1$&$-0.026~(-0.114, 0.059)$& $-1.572~(-1.714, -1.430)$&$0.735~(0.595, 0.879)$&$-1.545~(-1.745, -1.339)$\\
$\beta_2$&$0.892~(0.764, 1.018)$& $-0.225~(-0.326, -0.116)$&$0.159~(-0.032, 0.322)$&$-0.949~(-1.077, -0.821)$\\
$\beta_3$&$0.543~(0.493, 0.589)$& $-0.011~(-0.058, 0.040)$&$0.209~(0.162, 0.255)$&$0.195~(0.150, 0.238)$\\
$\beta_4$&$-0.461~(-0.511, -0.410)$& $0.089~(0.043, 0.134)$&$0.503~(0.452, 0.557)$&$0.052~(0.009, 0.095)$\\
$\beta_5$&$-3.593~(-3.661, -3.525)$& $-2.056~(-2.125, -1.982)$&$-1.192~(-1.254, -1.131)$&$-0.925~(-0.988, -0.863)$\\
$\beta_6$&$-2.626~(-2.686, -2.566)$& $-2.453~(-2.530, -2.378)$&$-2.545~(-2.609, -2.481)$&$-2.920~(-2.989, -2.857)$\\
$\beta_7$&$-7.286~(-7.404, -7.170)$& $-5.893~(-6.010, -5.778)$&$-4.970~(-5.072, -4.875)$&$-4.605~(-4.709, -4.507)$\\
$\beta_8$&$1.393~(1.248, 1.534)$&$1.050~(0.901, 1.187)$&$0.723~(0.598, 0.856)$&$0.645~(0.525, 0.759)$\\
\hline
$\lambda_1$&$0.590~(0.587, 0.592)$& $0.492~(0.489, 0.495)$&$0.691~(0.690, 0.694)$&$0.558~(0.555, 0.560)$\\
$\lambda_2$&$0.400~(0.398, 0.403)$& $0.473~(0.470, 0.476)$&$0.525~(0.523, 0.527)$&$0.441~(0.438, 0.443)$\\
$\lambda_3$&$0.010~(0.010, 0.011)$& $0.019~(0.019, 0.020)$&$-0.028~(-0.028, -0.027)$&$-0.013~(-0.014, -0.013)$\\
\hline
$\sigma^2_{\xi}$& $0.393~(\textnormal{fixed})$& $0.430~(\textnormal{fixed})$ & 0.341~(\textnormal{fixed})& $0.352~(\textnormal{fixed})$\\
\bottomrule
\end{tabular}}
\end{table}
}

\begin{figure}[ht!]
\centering
  \includegraphics[width=11cm, height=4.1cm]{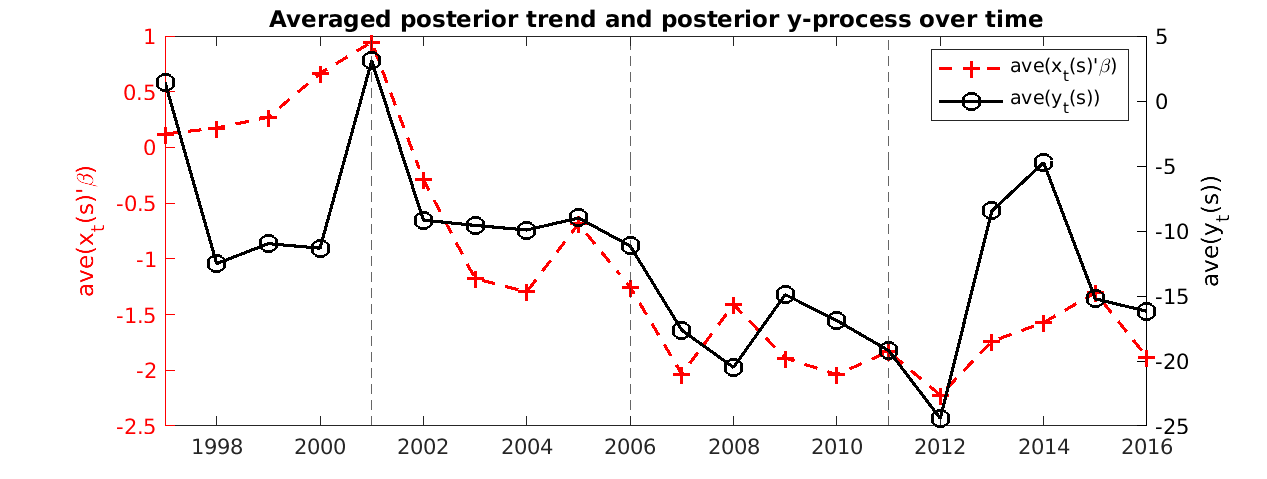}
\caption{The posterior means of $\textnormal{ave}\{\mb{x}_t(\mb{s})^{\prime}\sbf{\beta}: \mb{s}=(\textnormal{lon, lat}), \textnormal{and}\ \textnormal{lat}\in (\textnormal{lat}_0-\Delta, \textnormal{lat}_0+\Delta)\}$ and $\textnormal{ave}\{y_t(\mb{s}): \mb{s}=(\textnormal{lon, lat}), \textnormal{and}\ \textnormal{lat}\in (\textnormal{lat}_0-\Delta, \textnormal{lat}_0+\Delta)\}$, where the target latitude value is $\textnormal{lat}_0=75^{\circ}$N and $2\Delta=1^{\circ}$ is the bandwidth. The ends of Period 1, 2, 3, and 4 are 2001, 2006, 2011, and 2016, respectively, and are shown with a vertical dashed line.  }\label{Fig_RD_post_trend_y}
\end{figure}

Table~\ref{Table_RD_PE} gives the parameter-inference results for our proposed BHM, where both posterior means and corresponding $95\%$ credible intervals of parameters are reported. The covariates contribute in different ways, depending on the time period. To visualize the spatio-temporal variability that they capture, Figure~\ref{Fig_RD_post_trend_y} shows a time series plot of the posterior mean of $\textnormal{ave}\{\mb{x}_t(\mb{s})^{\prime}\sbf{\beta}\}$ as well as the corresponding time series plot of the posterior mean of $\textnormal{ave}\{y_t(\mb{s})\}$, where in each case $\mb{s}$ is averaged over a $1^{\circ}$ latitude band centered on $\textnormal{lat}_0=75^{\circ}$N. From the scales on the vertical axis in Figure~\ref{Fig_RD_post_trend_y}, the covariates are less important than the spatio-temporal variability in the random effects. 

\begin{figure}[ht!]
\centering
  \includegraphics[width=10cm, height=8.2cm]{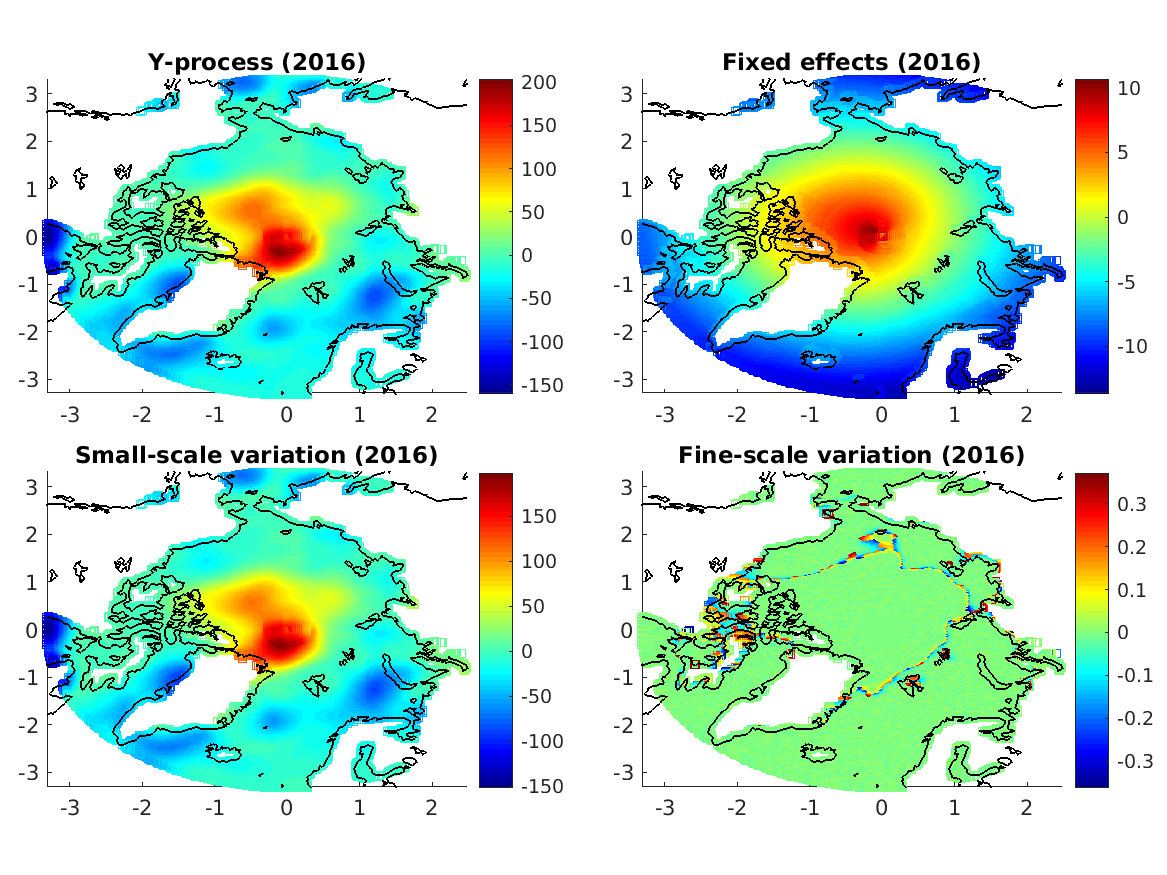}
\caption{Year $t=2016$. Top left: Posterior mean of the $y$-process. Top right: Posterior mean of the fixed-effects component of the $y$-process. Bottom left: Posterior mean of the small-scale component of the $y$-process. Bottom right: Posterior mean of the fine-scale component of the $y$-process. Adding these three components results in the posterior mean of the $y$-process, shown top left. (Note that the scales are different and the small-scale variation's contribution to the $y$-process dominates.)}\label{Fig_Y_all_component_1997}
\end{figure}

\begin{figure}[ht!]
\centering
\begin{subfigure}[$t=2001$, $y$-scale]{
  \includegraphics[width=10cm, height=4.2cm]{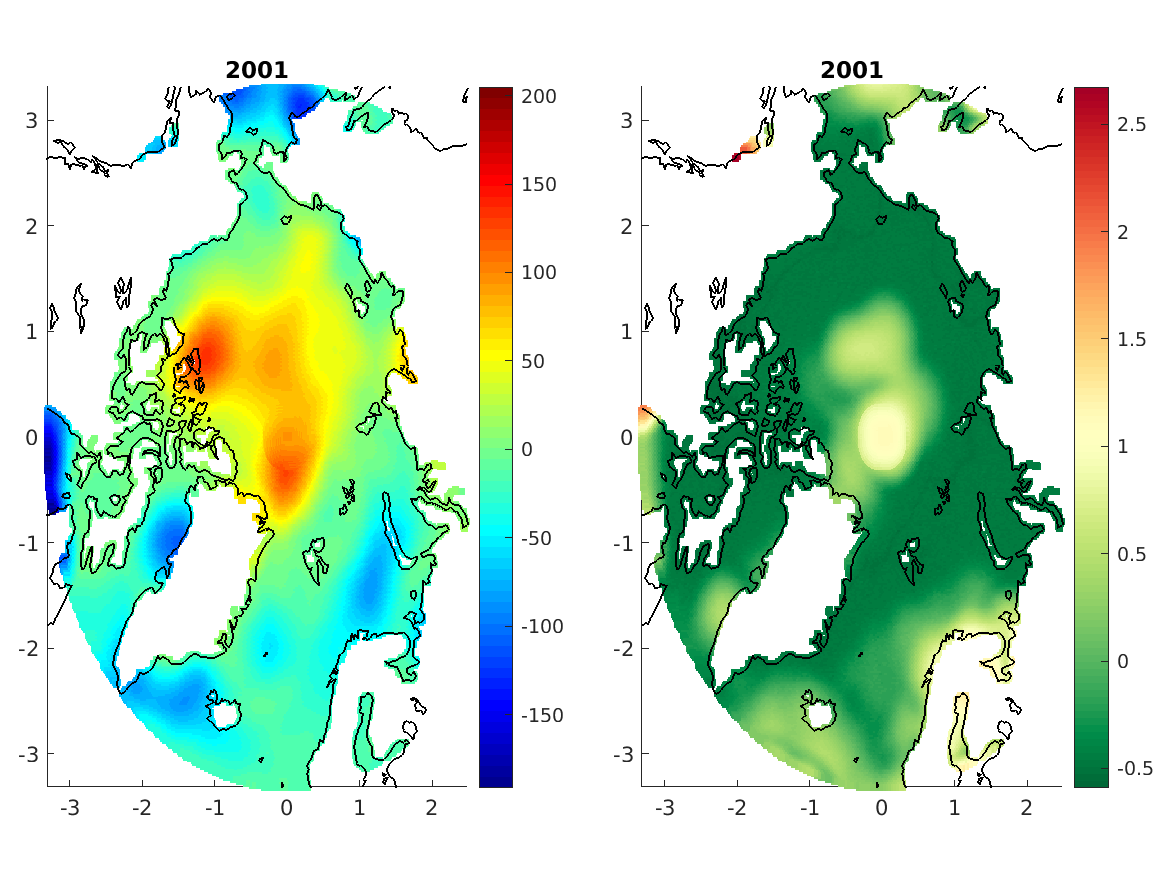}}
\end{subfigure}
\begin{subfigure}[$t=2016$, $y$-scale]{
  \includegraphics[width=10cm, height=4.2cm]{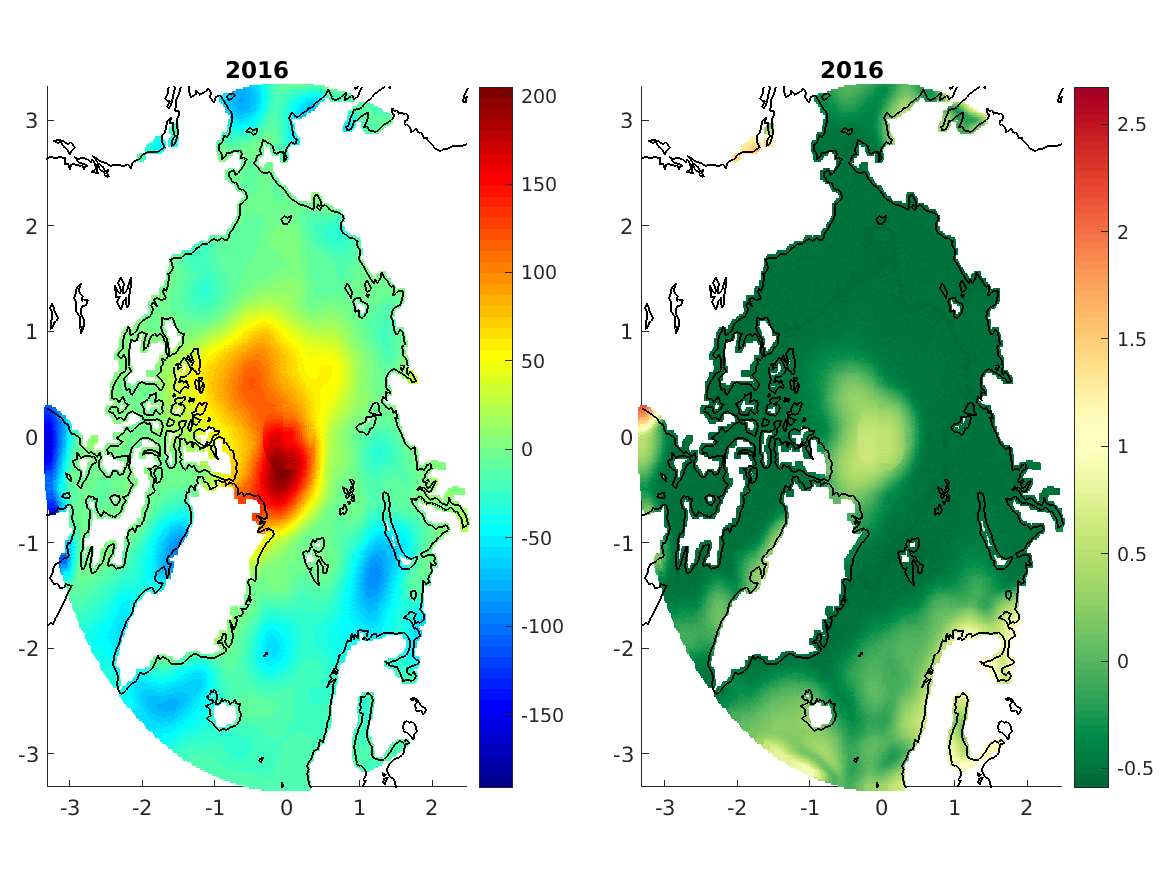}}
\end{subfigure}
\caption{The posterior means (left panels) and posterior standard deviations (on the log scale; right panels) of the logit-transformed process $\{y_t(\mb{s})\}$. (a) $t=2001$; (b) $t=2016$.  }\label{Fig_RD_Y_mean}
\end{figure}

\begin{figure}[ht!]
\centering
\begin{subfigure}[$t=2001$, posterior mean and s.d. ($p$-scale)]{
  \includegraphics[width=8cm, height=3.2cm]{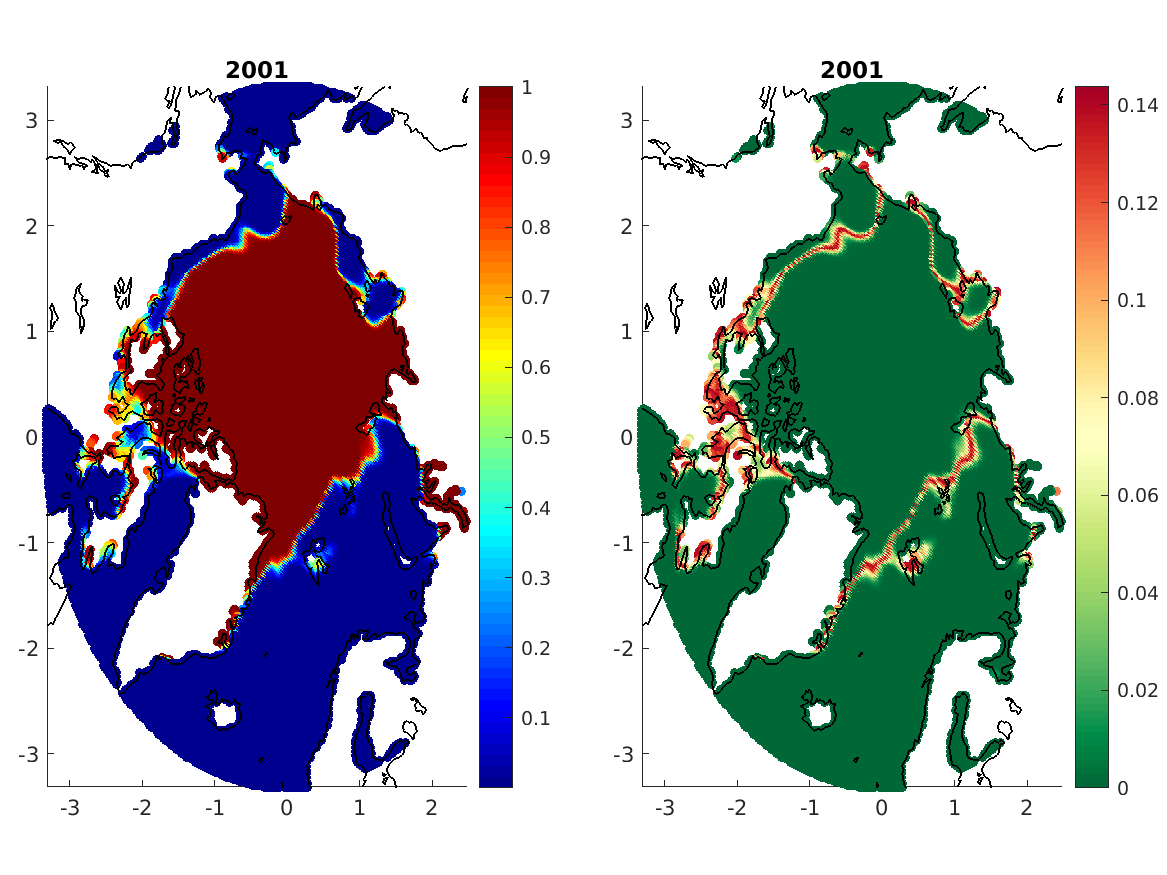}}
\end{subfigure}
\begin{subfigure}[$t=2001$, the $0.05$ and $0.95$ quantiles ($p$-scale)]{
  \includegraphics[width=8cm, height=3.2cm]{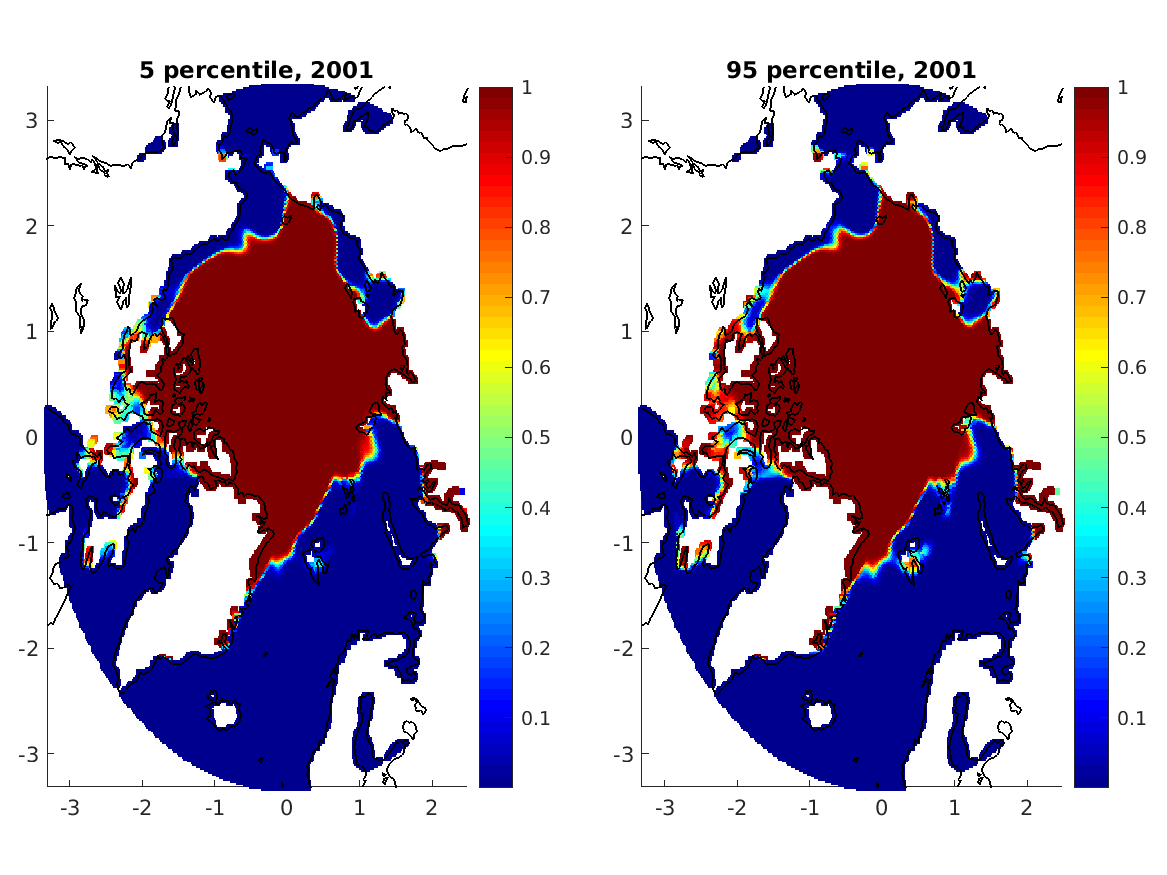}}
\end{subfigure}
\begin{subfigure}[$t=2016$, posterior mean and s.d. ($p$-scale)]{
  \includegraphics[width=8cm, height=3.2cm]{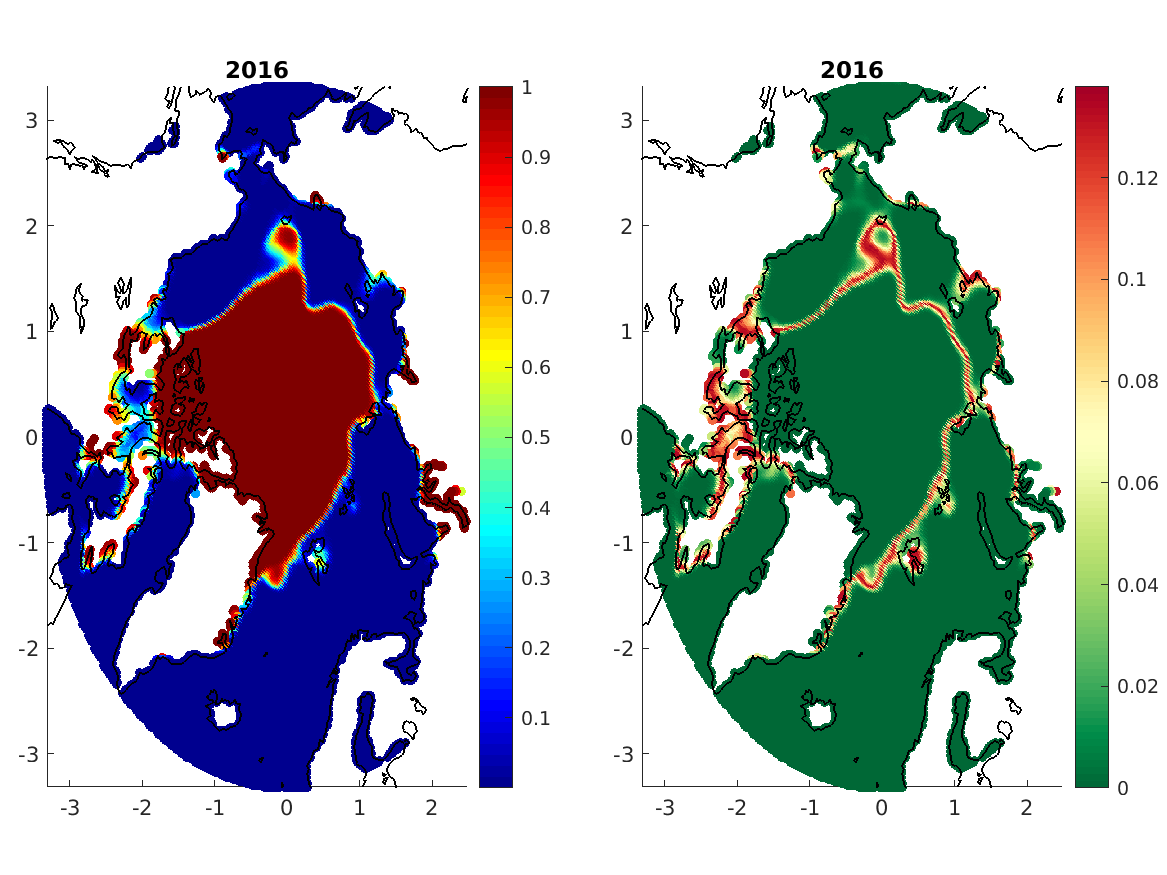}}
\end{subfigure}
\begin{subfigure}[$t=2016$, the $0.05$ and $0.95$ quantiles ($p$-scale)]{
  \includegraphics[width=8cm, height=3.2cm]{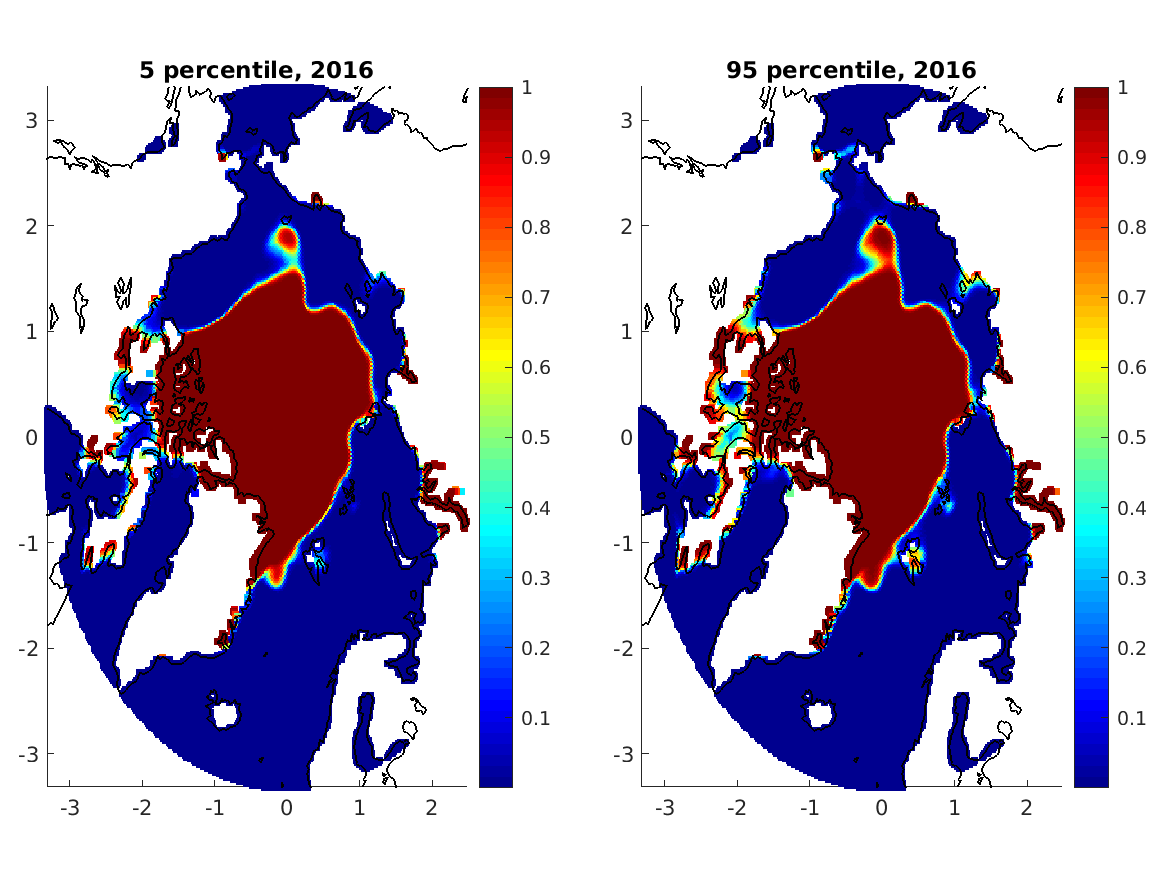}}
\end{subfigure}
\caption{(a) and (c): The posterior means (left panel) and posterior standard deviations (right panel) of the probability process $\{p_t(\mb{s})\}$. (b) and (d): Posterior $0.05$ quantile (left panel), and posterior $0.95$ quantile (right panel) of $p_t(\mb{s})$. The year $t=2001$ is shown in (a) and (b), and the year $t=2016$ is shown in  (c) and (d). }\label{Fig_RD_P_mean}
\end{figure}
For the propagator matrix $\mb{H}$, the within-resolution covariance parameters $\lambda_1$ and $\lambda_2$ have a dominant magnitude, and the correlations for Resolution-1 and Resolution-2 basis functions between $t$ and $t-1$ are always positive. The between-resolution covariance parameter $\lambda_3$ has a very small magnitude, where it is positive for the first two periods and then becomes negative for the latter two periods. The changing sign of $\lambda_3$ implies that the autoregressive structure of $\{\sbf{\eta}_t\}$ changes from the earlier decade to the recent decade, something we see in a striking way when we compare temporal semivariograms between the two decades (Section~\ref{Sec_RD_summaries}). 

Figure~\ref{Fig_Y_all_component_1997} shows the posterior mean of the latent process $\{y_{2016}(\mb{s})\}$, along with its fixed-effects and its random-effects components. The detailed spatial variations are mainly captured by the small-scale-variation component, $\{\mb{S}(\mb{s})^{\prime}\sbf{\eta}_{2016}\}$, while the amplitude of the fine-scale-variation component is tiny and mostly located around the ice\textendash water boundaries. 

Figure~\ref{Fig_RD_Y_mean} and Figure~\ref{Fig_RD_P_mean} show the posterior means and posterior standard deviations of $\{y_t(\mb{s}):t=2001, 2006\}$ and $\{p_t(\mb{s}):t=2001, 2006\}$, respectively; the $0.05$ and $0.95$ posterior quantiles of $\{p_t(\mb{s})\}$ are also displayed. We observe that on the logit scale ($y$-scale), large uncertainties typically appear at locations where process values have large magnitudes. On the probability scale ($p$-scale), large uncertainties occur at the ice\textendash water boundary locations, as expected. It is clear that the Arctic sea ice has decreased substantially from the end year of Period 1 (2001) to the end year of Period 4 (2016).

\subsection{Evolution of Arctic sea ice} \label{Sec_RD_summaries}

\begin{figure}[ht!]
\centering
  \includegraphics[width=10cm, height=10cm]{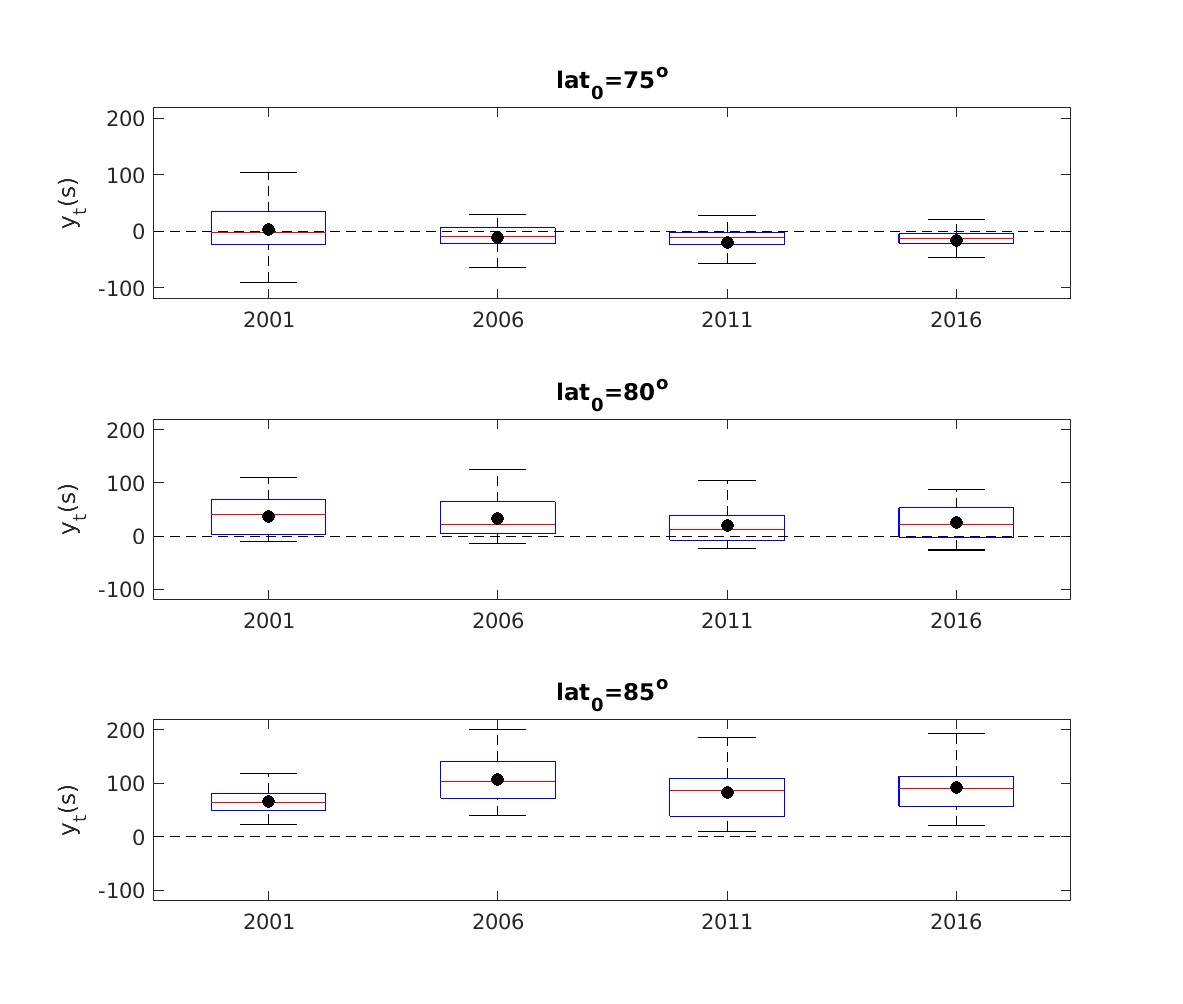}
\caption{Boxplots of $\{E(y_t(\mb{s})|\mb{Z}^o): \mb{s}=(\textnormal{lon, lat}), \textnormal{and}\ \textnormal{lat}\in (\textnormal{lat}_0-\Delta, \textnormal{lat}_0+\Delta)\}$ for three latitude bands and for years $t=2001, 2006, 2011,2016$, where the width of the latitude band is $2\Delta=1^{\circ}$. The dashed line indicates the value of zero for $y_t(\mb{s})$ (which corresponds to $0.5$ on the probability scale).} \label{Fig_RD_Boxplots}
\end{figure}

In this subsection, we first consider the spatial variability of the latent process (on the $y$-scale) in different latitude bands for the end years of Periods~1~\textendash~4. Specifically, at $t=2001,2006,2011,2016$, we compare the five-number boxplots of $\{E(y_t(\mb{s})|\mb{Z}^o)\}$ within the three latitude bands around $75^{\circ}$N,  $80^{\circ}$N, and $85^{\circ}$N. Figure~\ref{Fig_RD_Boxplots} shows the boxplots augmented with the average (denoted by a dot); by varying $\mb{s}=(\textnormal{lon, lat})$ within the given latitude band $(\textnormal{lat}_0-\Delta, \textnormal{lat}_0+\Delta)$ for $2\Delta=1^{\circ}$, we obtain the boxplot. Because on the $y$-scale, the value $0$ corresponds to $0.5$ on the probability scale, which is a natural cut-off value that says the potential for ice and water is the same, we feature the value $0$ in the boxplots on the $y$-scale. By looking across the columns, we see the evolution of Arctic sea ice over the two decades of data, at time points five years apart. The earlier decade is characterized by the behavior at $t=2001$ and $2006$, and the recent decade by the behavior at $t=2011$ and $2016$. 

The evolution is clear at $\textnormal{lat}_0=75^{\circ}$N and $80^{\circ}$N: The central tendencies decrease over time (more water and less ice). At $85^{\circ}$N, the sea ice appears to be in equilibrium over the period of observation, but a more refined (temporal semivariogram) analysis below, shows oscillatory behavior in the recent decade. 

Figure~\ref{Fig_bp_pscale} shows similar boxplots on the $p$-scale in different latitude bands, which provide a clearer contrast between the earlier and the latest decade. A detailed discussion is given in Section~1, where attention is directed towards a possible future collapse of the sea ice at latitude $80^{\circ}$N.
\begin{figure}[ht!]
\centering
\begin{subfigure}[$y$-scale]{
  \includegraphics[width=5.5cm, height=5cm]{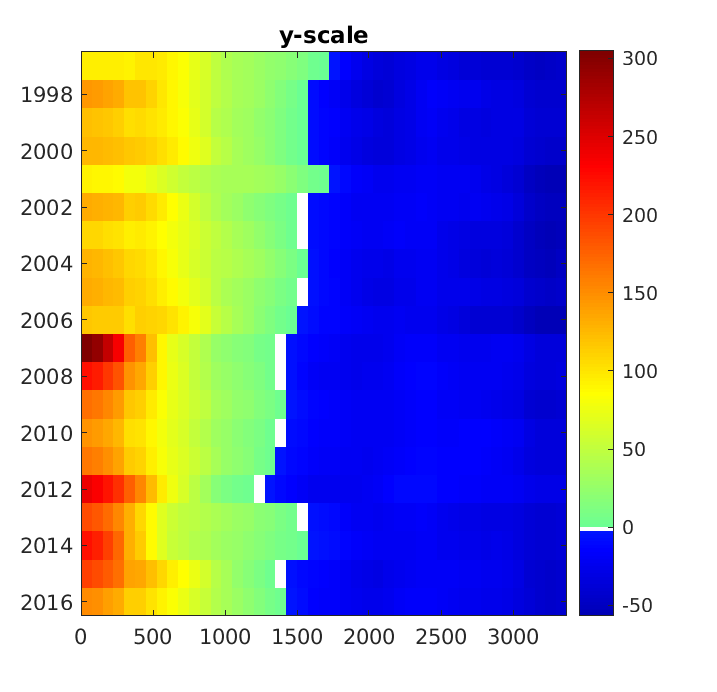}}
\end{subfigure}
\begin{subfigure}[$p$-scale]{
  \includegraphics[width=5.5cm, height=5cm]{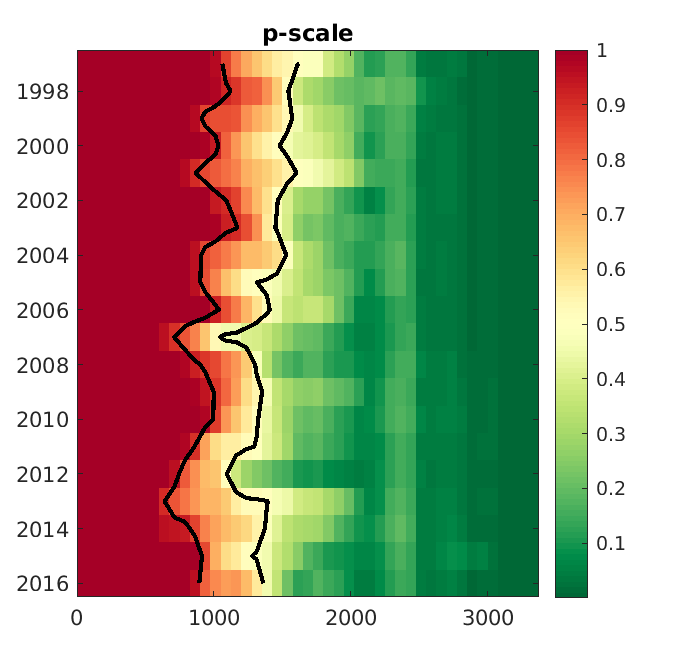}}
\end{subfigure}
\caption{Hovm\"{o}ller diagrams showing the posterior means of the spatial averages, $\{\bar{y}_t(x)\}$ (left panel) and $\{\bar{p}_t(x)\}$ (right panel), for each year, where the average is taken over all pixels at distance $x\pm h$ from the North Pole. The horizontal axis is in units of km, and the vertical axis is in units of year. The right panel also shows the $0.9$ and $0.5$ contours from left to right. }\label{Fig_RD_Hov_global}
\end{figure}

The yearly temporal dynamics of the latent process $\{y_t(\mb{s})\}$ (as well as that of $\{p_t(\mb{s})\}$) can be visualized through Hovm\"{o}ller diagrams \citep[e.g.,][Section~5.1.2]{cressie2015statistics}. Figure~\ref{Fig_RD_Hov_global} shows the Hovm\"{o}ller diagrams of the posterior means of both $\{y_t(\mb{s})\}$ and $\{p_t(\mb{s})\}$, where the horizontal axis is the great-circle distance of $\mb{s}$ to the North Pole in units of km, and the vertical axis is time in units of year. Specifically, for each distance $x$ and year $t$, we obtained the spatially averaged posterior mean of $y_t(\mb{s})$ as $\bar{y}_t(x)=\textnormal{ave}\{E(y_t(\mb{s})|\mb{Z}^o): \|\mb{s}-\mb{s}^*\|\in (x-h, x+h)\}$, where $\mb{s}^*$ is the North Pole, $2h=75\textnormal{km}$ is a \textit{distance bandwidth}, and $\|\mb{s}-\mb{s}^*\|$ computes the great-circle distance between two spatial locations. An analogous quantity on the $p$-scale, denoted by $\bar{p}_t(x)$, is obtained and can be interpreted as the potential for sea ice at $(x,t)$. These are plotted as a function of $(x, t)$ in Figure~\ref{Fig_RD_Hov_global}. 

The $0.9$ and $0.5$ contours plotted on the $p$-scale show that in the second decade the potential for sea ice at distances $1000$~km or more has decreased substantially. Section~S2 of the Supplementary Material shows, through Hovm\"{o}ller diagrams in different regions of the Arctic, that this decrease is not uniform. The regional Hovm\"{o}ller diagrams in Section~S2 provide a very effective way of assessing where the sea ice is most sensitive to a warming climate. 

\begin{figure}[ht!]
\centering
\begin{subfigure}[$\textnormal{lat}_0=75^{\circ}$N]{
  \includegraphics[width=12cm, height=3cm]{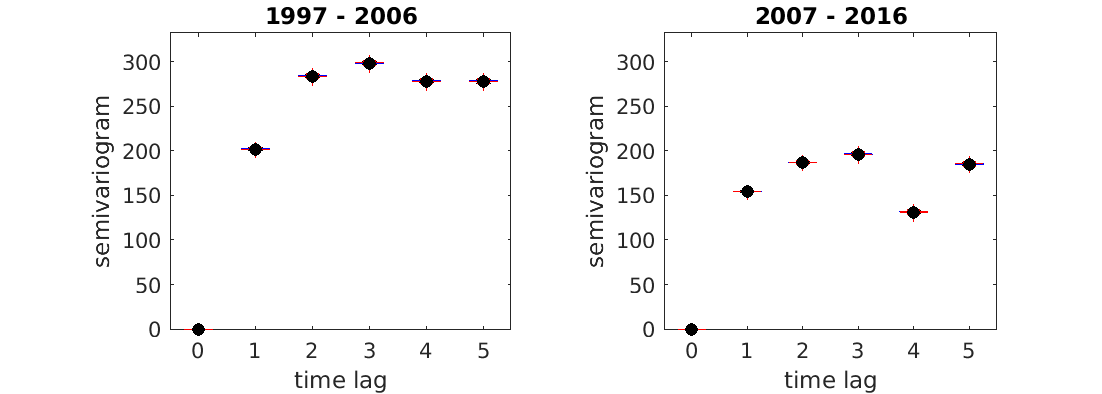}}
\end{subfigure}
\begin{subfigure}[$\textnormal{lat}_0=80^{\circ}$N]{
  \includegraphics[width=12cm, height=3cm]{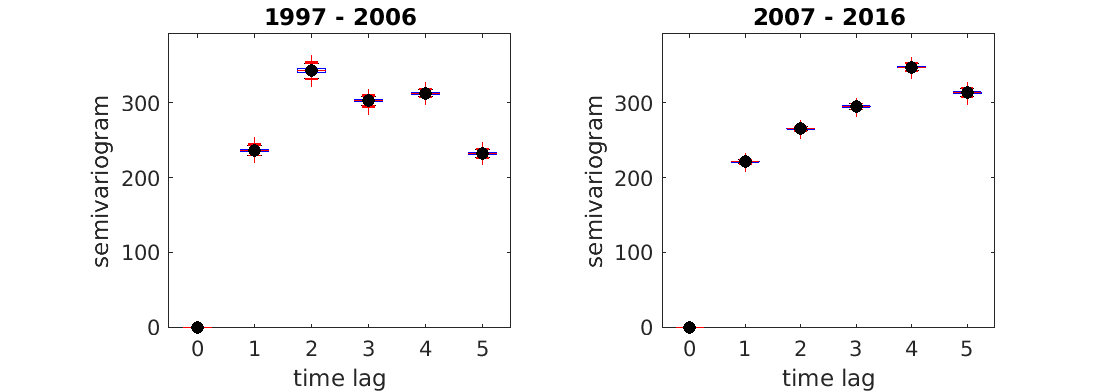}}
\end{subfigure}
\begin{subfigure}[$\textnormal{lat}_0=85^{\circ}$N]{
  \includegraphics[width=12cm, height=3cm]{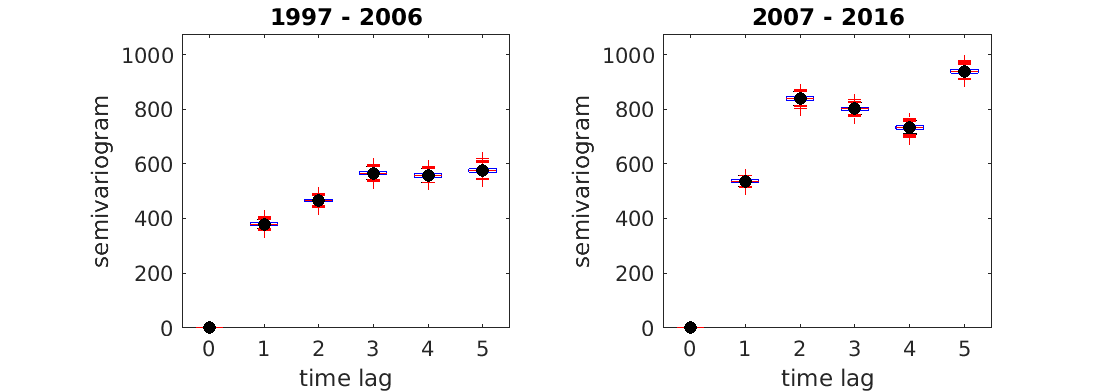}}
\end{subfigure}
\caption{The boxplots and the average (indicated by the solid black dot) of samples from the posterior distribution of temporal semivariograms, $\{\bar{\gamma}(h;\textnormal{lat}_0): h=1,\ldots, 5\}$, for the three latitude bands, around $75^{\circ}$N, $80^{\circ}$N, and $85^{\circ}$N, where the width of the latitude band is $2\Delta=1^{\circ}$. }\label{Fig_RD_Temporal_semivariograms}
\end{figure}

The sea-ice extent plotted in Figure~\ref{Fig_data_20years} and the potential for sea ice ($p$-scale) plotted in Figure~\ref{Fig_RD_Hov_global}(b) indicate a slower and deeper oscillatory behavior in the recent decade. We investigate the temporal-dependence structure for the two decades by using the empirical temporal semivariogram computed for the detrended latent process for different latitude bands: For $h=1,\ldots, M$,
\begin{eqnarray*}
\bar{\gamma}(h;\textnormal{lat}_0)\equiv\frac{1}{2}\textnormal{ave}\{(\delta_t(\mb{s})-\delta_{t+h}(\mb{s}))^2: \mb{s}=(\textnormal{lon, lat}), \textnormal{and}\ \textnormal{lat}\in  (\textnormal{lat}_0-\Delta, \textnormal{lat}_0+\Delta)\}, 
\end{eqnarray*}
where $\delta_t(\mb{s})\equiv \mb{S}_t(\mb{s})^{\prime}\sbf{\eta}_t+\xi_t(\mb{s})$ is the random-effect component obtained from a sample from the posterior distribution, $\mb{s}\equiv (\textnormal{lon, lat})^{\prime}$, $\textnormal{lat}_0$ is the target latitude, $2\Delta=1^{\circ}$ is the width of the latitude band, and the upper temporal lag, $M$, is half the range of possible lags rounded up \citep[e.g.,][]{Cressie1988A}. Then we obtained the five-number boxplots of the posterior distribution of $\bar{\gamma}(h;\textnormal{lat}_0)$ as well as $E(\bar{\gamma}(h;\textnormal{lat}_0)|\mb{Z}^o)$ (denoted by a dot) from the MCMC samples from the posterior distribution of $\{\delta_t(\mb{s})\}$ for the earlier decade, 1997~\textendash~2006, and for the recent decade, 2007~\textendash~2016, where in both cases $M=\lfloor 9/2\rfloor=5$. Figure~\ref{Fig_RD_Temporal_semivariograms} shows these for $\textnormal{lat}_0=75^{\circ}\textnormal{N}, 80^{\circ}\textnormal{N}$, and $85^{\circ}\textnormal{N}$, and it should be noted from the boxplots that our posterior inference is very precise. There are several remarkable features that indicate differences between the two decades. At $\textnormal{lat}_0=75^{\circ}$N, the sill is much smaller for the recent decade, yet the sill is much larger for the recent decade at $\textnormal{lat}_0=85^{\circ}$N (and about the same at $\textnormal{lat}_0=80^{\circ}$N). Thus, the contrast in the evolution of the sea ice has flipped when comparing higher latitudes to lower latitudes. In the recent decade, the consistent ``dip" at lag~4 of the semivariograms for $\textnormal{lat}_0=75^{\circ}$N and $\textnormal{lat}_0=85^{\circ}$N indicates a periodicity that is reflected in the substantial oscillations around the trend line seen in Figure~\ref{Fig_data_20years}. That it appears at high latitudes indicates the need to monitor the sea ice for evolutionary changes both away from and close to the North Pole. 

The empirical semivariograms have added to our understanding of the temporal behavior of Arctic sea ice, already obtained from boxplots and Hovm\"{o}ller diagrams. Of concern is a possible scenario in the future where the polar ice cap collapses in a matter of years not decades. 

\section{Conclusion}

In this article, we develop a Bayesian hierarchical model (BHM) for the binary Arctic sea-ice-extent (SIE) data, where the $0$\textendash$1$ observations are modeled using a logistic link function in the data model and a spatio-temporal linear mixed-effects model in the process model. Compared with the empirical hierarchical model (EHM) in \cite{zhang2018estimating}, the BHM accounts for parameter uncertainties and, in Section~S1 of the Supplementary Material, we show in a simulation study that a BHM can yield better prediction accuracy than an EHM in spatial gaps where there are no nearby data. Furthermore, the BHM allows the uncertainties of parameters such as regression coefficients and covariance parameters to be directly obtained from MCMC samples, which allows proper statistical interpretation of the effects of covariates and the data-dependence structure. In contrast, parameter uncertainties from the EM algorithm require numerical differentiations \citep[e.g.,][]{meng1991using,jamshidian2000standard}. When constructing the ST-GLMM for the latent process, several physically motivated covariates were used for the fixed-effects component. The BHM yields credible intervals for the coefficients of these covariates, and they are almost all substantially different from zero. 

When modeling the Arctic SIE data, we assume $\mb{H}_t$ stays the same in each five\textendash year time period. This can be extended to the case where $\mb{H}_t$ varies over time by putting a prior on the distribution of $\mb{H}_t$ \citep[e.g.,][]{berliner2000long}. Further parameterization of $\mb{K}$ increases the degrees of the freedom and would lead to better estimation of both $\mb{K}$ and $\sigma^2_{\xi}$. For example, $\mb{K}$ may be parameterized using a parametric covariance function, where the basis-function centers are the spatial locations \citep[e.g.,][]{katzfuss2009maximum}. Alternatively, the neighborhood information of the basis-function centers may be used to form a known baseline covariance matrix $\mb{K}_0$ \citep{bradley2015multivariate} or a known baseline precision matrix $\mb{Q}_0$ \citep[e.g.,][]{reich2006effects,hughes2013dimension}. Then let $\mb{K}=\sigma^2_{K}\mb{K}_0$ or $\mb{K}^{-1}=\tau^2_K\mb{Q}_0$, which involves only one unknown parameter $\sigma^2_K$ (or $\tau^2_K$) to be estimated.

The focus of our paper is to propose useful functionals that characterize the space-time changes of Arctic sea ice, particularly between an earlier decade and a recent decade. The boxplots of the posterior means of $\{p_t(\mb{s})\}$ in Figure~\ref{Fig_bp_pscale} in intermediate latitude bands show an obvious declining tendency over two decades. We also used Hovm\"{o}ller diagrams to visualize the temporal dynamics of the posterior means of the latent processes, $\{y_t(\mb{s})\}$ and $\{p_t(\mb{s})\}$, both for the whole Arctic region (Section~\ref{Sec_RD_summaries}) and regionally (Section~S2 of the Supplementary Material). We found that for the entire Arctic, the 0.5-probability contour of the posterior mean of the latent probability process, $\{p_t(\mb{s})\}$, which represents an important boundary of the sea\textendash ice potential, has moved northwards over time. Not surprisingly, this behavior is more obvious for years when substantially smaller Arctic SIE was observed (e.g., 2007 and 2012). Furthermore, we made a comparison of the empirical temporal semivariograms between the earlier decade before 2007 and the recent decade from 2007 at different latitudes, and we found that their temporal-dependence structures were very different. Our results demonstrate the importance of monitoring sea ice at high latitudes as well as at latitudes where the sea\textendash ice boundary occurs, and they propose summaries of the data that detect changes in the sea ice. We have also observed and summarized highly oscillatory behavior in the recent decade, noting the possibility of substantial changes in a matter of years rather than decades.

A change in Arctic sea ice can impact the weather in other regions, which indicates that future research should focus on teleconncections of the effects of Arctic forcing on other environmental variables (e.g., surface temperature and precipitation) in other regions \citep{cohen2014recent}. Another important research project would be the albedo\textendash ice feedback of the Arctic region, which could be studied using a bivariate spatio-temporal statistical model of sea-ice concentrations and albedo. 


\section*{Supplementary Material}
The Supplementary Material contains a simulation study that compares the inference performance of the empirical hierarchical model (EHM) and the Bayesian hierarchical model (BHM) in Section~S1; a detailed discussion of the dynamics of sea ice in different regions of the Arctic in Section~S2; details of the MCMC sampling algorithm and convergence diagnostics for the Arctic sea-ice-extent data in Section~S3; the classification accuracy of the Bayesian inference in Section~S4; and visualization of the ice-to-water and water-to-ice transition probabilities in Section~S5.

\section*{Acknowledgement}
Cressie's research is supported by an Australian Research Council Discovery Project, number DP190100180. Zhang's research is supported by the Fundamental Research Funds for the Central Universities, China, and National Science Foundation China Project, number 11901316. It is also partially supported by the Key Laboratory of Pure Mathematics and Combinatorics (LPMC) of Ministry of Education, China and the Key Laboratory for Medical Data Analysis and Statistical Research (KLMDASR) of Tianjin, China. We would like to thank Bruno Sans\'{o} for encouraging us to write up our research into this article, and we are very grateful for the many comments and suggestions from the Associate Editor and four reviewers. 

\bibliographystyle{chicago}
\bibliography{ref-ST-GLM}

\end{document}


\renewcommand{\baselinestretch}{1.2}

\markright{ \hbox{\footnotesize\rm Supplement
}\hfill\\[-13pt]
\hbox{\footnotesize\rm
}\hfill }

\renewcommand{\thefootnote}{}
$\ $\par \fontsize{11}{14pt plus.8pt minus .6pt}\selectfont


\centerline{\large\bf Bayesian spatio-temporal modeling of Arctic sea ice extent}
\vspace{.25cm}
 \centerline{Bohai Zhang$^*$ and Noel Cressie$^\dag$}
\vspace{.4cm}
 \centerline{\it Nankai University$^*$ and University of Wollongong$^\dag$}
\vspace{.55cm}
 \centerline{\bf Supplementary Material}
\vspace{.55cm}
\fontsize{9}{11.5pt plus.8pt minus .6pt}\selectfont

\noindent

The Supplementary Material contains a simulation study that compares the inference performance of the empirical hierarchical model (EHM) and the Bayesian hierarchical model (BHM) in Section~\ref{Sec_sim_comp}; a detailed discussion of the dynamics of sea ice in different regions of the Arctic in Section~\ref{Sec_Append_Hov_local}; details of the MCMC sampling algorithm and convergence diagnostics for the Arctic sea-ice-extent data in Section~\ref{Sec_algorithm_diagnostics}; the classification accuracy of the Bayesian inference in Section~\ref{Sec_classification}; and visualization of the ice-to-water and water-to-ice transition probabilities in Section~\ref{Sec_IWT_WIT}.

\par

\setcounter{section}{0}
\setcounter{equation}{0}

\def\theequation{S\arabic{section}.\arabic{equation}}
\def\thesection{S\arabic{section}}

\def\thefigure{S\arabic{figure}}
\def\thetable{S\arabic{table}}

\fontsize{11}{14pt plus.8pt minus .6pt}\selectfont

\section{Simulation studies and validation comparison of the EHM and the BHM} \label{Sec_sim_comp}

In this section, we compare the prediction performances of the EHM and the BHM (with the fine-scale-variation variance $\sigma^2_{\xi}$ estimated) through a simulation study. We employed a similar simulation design to that in \cite{zhang2018estimating}, but now we consider both parameter inference and prediction performance. The spatial domain is on the unit square, $[0, 1]\times [0,1]$, consisting of $N=100^2=10,000$ regular-grid locations $\mathcal{D}\equiv\{\mb{s}_i=(s_{i,1},s_{i,2}):i=1,\ldots,N\}$, for each of $T=6$ time points. We then partitioned the data into a training data set and a validation data set, where the validation data set was made up of a prediction data set and a forecasting data set. For $T=1,\ldots, 5$, we held out the spatial locations in the rectangular domain, $[0.7, 1]\times [0.8, 1]$, which define part of the prediction data set; these $600$ locations represent a missing-by-design (or MBD) scenario. The other part of the prediction data set was defined by randomly selecting $600$ spatial locations from the remaining locations, which represents a missing-at-random (or MAR) scenario. Combined, the two scenarios define missing-data locations $\{\mb{s}_{t,j}^u\}$. The remaining $8800$ locations $\{\mb{s}_{t,i}^o\}$ are used as the training data set, for $t=1,\ldots, 5$. For the forecasting data set, all $10,000$ locations at $t=6$ were predicted. 

\begin{figure}[ht!]
\centering
  \includegraphics[width=8cm, height=6cm]{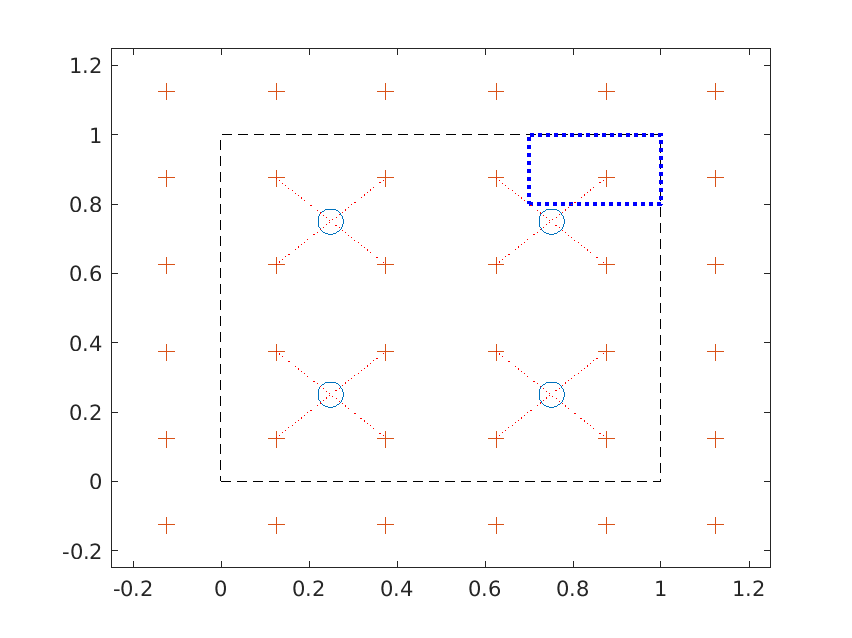}
 \caption{Centers of bisquare basis functions, where circles and pluses are for Resolution-1 and Resolution-2 basis functions, respectively. The extent of the spatial domain $\mathcal{D}\equiv[0,1]\times[0,1]$ is indicated by the dashed lines. The dotted lines connect Resolution-2 centers in $\mathcal{D}$ to their neighboring Resolution-1 centers. The bold dotted lines in the top-right corner indicate the hold-out rectangular region used for prediction. }\label{Fig_basis_center}
\end{figure} 

We simulated a data set from the hierarchical statistical model given by (1) and (2) in the main paper. 
Along with the intercept, we used as covariate the distance to the domain center, $(0.5, 0.5)$, and we set $\sbf{\beta}=(5, -15)^{\prime}$. To generate the latent process, we specified the basis-function matrix $\mb{S}$, the matrix parameters $\mb{K}$, $\mb{H}$, $\mb{U}$, and the fine-scale-variation variance $\sigma^2_{\xi}$ as follows: We considered a two-resolution design for the basis functions, where Resolution 1 contains $r_1=2\times2=4$ basis functions, and Resolution 2 contains $r_2=6\times6=36$ basis functions; see Figure~\ref{Fig_basis_center}. Hence, in total there are $r=r_1+r_2=40$ spatial basis functions. Note that some Resolution-2 basis-function centers are outside the study domain to account for boundary effects \citep{cressie2010high}. Then each basis function was standardized by subtracting its sample mean and dividing by its sample standard deviation. 

Following \cite{cressie2008frk}, $\mb{K}\equiv\mathrm{var}(\sbf{\eta}_1)$ was specified such that $\mb{SKS}^{\prime}$ approximates a target covariance matrix $\sbf{\Sigma}_0$, where $\sbf{\Sigma}_0$ was obtained from an exponential covariance function, $\sigma^2\exp(-h/\psi)$, for $\sigma^2=1$ and $\psi=0.2$ \citep[practical range $=3\psi=0.6$; e.g.,][]{Bevilacqua:2012}. We first obtained $\mb{K}_0$ such that it minimizes the distance to $\sbf{\Sigma}_0$ in terms of the Frobenius norm. Then $\mb{K}=0.95(\mb{K}_0/(\mathrm{trace}(\mb{SK}_0\mb{S}^\prime)/N))$, in order to have $95\%$ of the total variation due to the small-scale random effects $\sbf{\eta}$. Then the fine-scale-variation variance $\sigma^2_{\xi}$ is set equal to $0.05$, to make the total variation equal to $1$. For $\mb{H}$ given by (4), 
$\mb{R}$ is the $0$\textendash$1$ adjacency matrix where each Resolution-1 (coarse resolution) basis-function center is connected to its nearest four Resolution-2 (fine resolution) neighbors, and we specified $\rho_1=0.4$, $\rho_2=0.4$, and $\rho_3=0.035$. Then the innovation matrix is $\mb{U}=\mb{K}-\mb{HKH}^\prime$, which is positive-definite for the $\mb{K}$ and $\mb{H}$ specified above. 

{\begin{table}[!ht]
\caption{Parameter-inference results for the BHM (with plug-in EM estimate $\hat{\sigma}^2_{\xi}$) and the EHM. For the EHM, the EM estimates of model parameters are reported while for the BHM, the posterior mean and the $95\%$ credible interval of parameters (in parentheses) are reported. }
\label{Table_Sim_PE}
\centering
\vspace{8pt}
\scalebox{0.8}{\begin{tabular}{|c|c|c|}
\toprule
Parameter& BHM (plug-in $\hat{\sigma}^2_{\xi})$& EHM\\
\hline
$\beta_{0}~(5)$&$5.045~(4.923,5.169)$ & $5.517$\\
$\beta_1~(-15)$& $-15.184~(-15.518,-14.851)$& $-16.497$\\
$\sigma^2_{\xi}~(0.05)$&  $0.050~(\textnormal{fixed})$& $0.050$\\
$\lambda_1~(0.4)$& $0.244~(-0.062,0.537)$& $0.282$\\
$\lambda_2~(0.4)$ & $0.307~(0.078,0.528)$& $0.409$\\
$\lambda_3~(0.035)$&$0.046~(-0.004,0.097)$ & $0.055$\\
\bottomrule
\end{tabular}}
\end{table}
}

We compared the BHM to the EHM, first in terms of inference on the parameters. For the EHM, we ran the EM algorithm given in \cite{zhang2018estimating} to obtain EM estimates of $\sbf{\theta}$. For the BHM, we put prior distributions on $\sbf{\theta}$ but fixed $\sigma^2_{\xi}$ at its EM estimate, and then we obtained joint posterior samples of $\sbf{\theta}\textbackslash\{\sigma^2_{\xi}\}$ and of the random effects $\sbf{\eta}$ and $\sbf{\xi}^o$. The inference results for the scalar parameters, along with the parameters' true values, are given in Table~\ref{Table_Sim_PE}. Compared to the EHM, we observe that the BHM gives much better parameter-inference results for the regression coefficients $\sbf{\beta}$, and the BHM's $95\%$ credible intervals of $\sbf{\beta}$, $\lambda_1$, $\lambda_2$, and $\lambda_3$ cover their respective true values, although the widths of the $95\%$ credible intervals for $\{\lambda_1, \lambda_2, \lambda_3\}$ are relatively large. In the EHM, the EM estimate of $\sbf{\beta}$ based on a single realization of the data shows a relatively large bias for $\sbf{\beta}$, which can affect the prediction performance of the EHM. 

{\begin{table}[!ht]
\caption{Validation results for the BHM (with plug-in EM estimate $\hat{\sigma}^2_{\xi}$) and the EHM (EM and TRUE), where root-mean-squared prediction errors (RMSPE) are reported for each of the missing-by-design (MBD) and missing-at-random (MAR) scenarios.}
\label{Table_Sim_pred}
\centering
\vspace{8pt}
\scalebox{0.8}{\begin{tabular}{|c|cc|cc|cc|}
\toprule
Method& \multicolumn{2}{c}{$y$-scale}& \multicolumn{2}{c}{$p$-scale}& \multicolumn{2}{|c|}{Forecasting}\\
\cline{2-7}
& MBD& MAR & MBD& MAR & $y$-scale& $p$-scale \\
\hline
BHM (fixed $\sigma^2_{\xi}$)& $0.50$& $0.31$& $0.04$& $0.04$& $1.21$&$0.15$\\
EHM-EM & $1.32$& $0.45$& $0.07$& $0.04$& $1.39$&$0.17$\\
EHM-TRUE& $0.36$& $0.29$& $0.03$& $0.04$&$1.12$&$0.16$\\
\bottomrule
\end{tabular}}
\end{table}
}

\begin{figure}[ht!]
\centering
\begin{subfigure}[$t=1$]{
  \includegraphics[width=13cm, height=3.5cm]{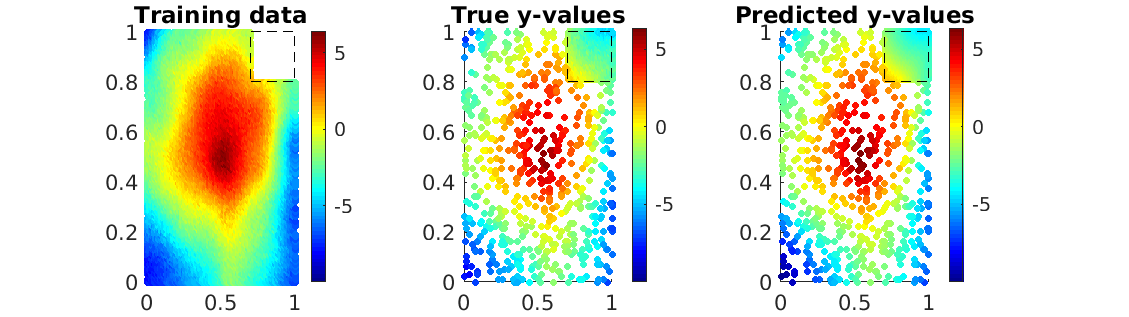}}
\end{subfigure}
\vspace{8pt}
\begin{subfigure}[$t=3$]{
  \includegraphics[width=13cm, height=3.5cm]{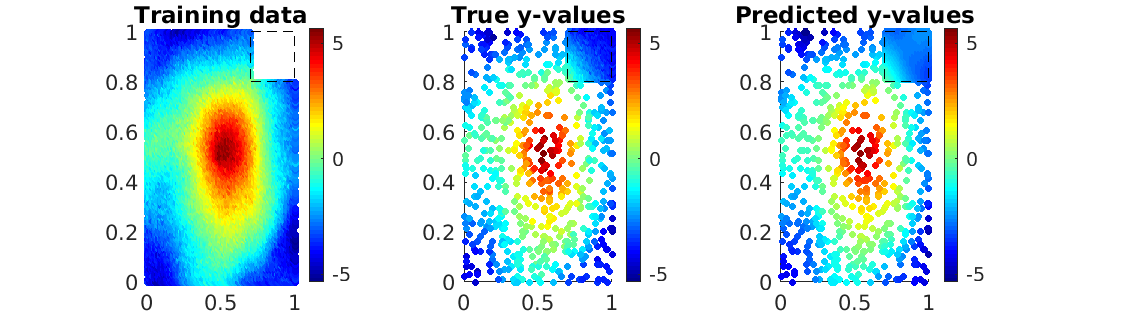}}
\end{subfigure}
\vspace{8pt}
\begin{subfigure}[$t=6$]{
  \includegraphics[width=13cm, height=3.5cm]{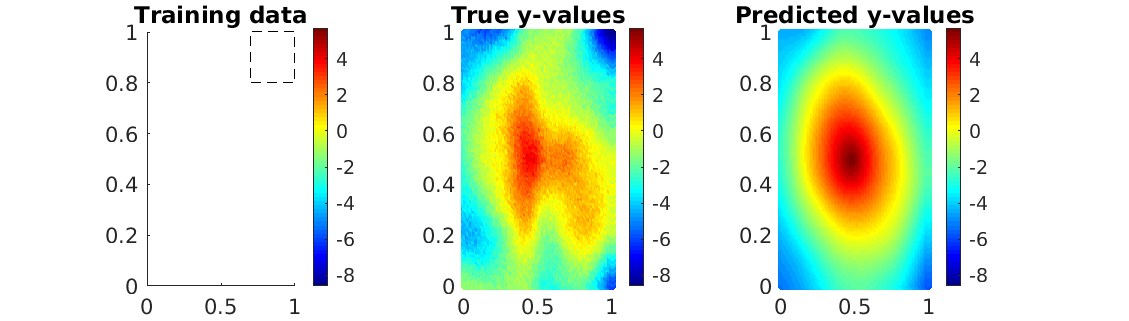}}
\end{subfigure}
\caption{Left panels: The training data $\{y_t(\mb{s}_i^o)\}$ for parameter estimation and prediction; middle vertical panels: The true values of $\{y_t(\mb{s}_j^u)\}$ used to calculate RMSPEs; right panels: The corresponding predictive means of $\{y_t(\mb{s}_j^u)\}$ using the BHM. The results are shown for $t=1,3,6$, where $t=6$ corresponds to the forecasting case. The dashed lines indicate the hold-out rectangular region, $[0.7, 1]\times [0.8, 1]$, for prediction. }\label{Fig_Sim_Pred_Y}
\end{figure}

For the EHM, the EM estimates and the true values of $\sbf{\theta}$ were separately substituted into the hierarchical model given by (1) and (2), 
from which predictive samples of $\sbf{\eta}$ and $\sbf{\xi}^o$ were generated, which in turn resulted in the predictions and forecasts. For the BHM, posterior samples for prediction and forecasting were obtained along the lines described in Section~3.2. Table~\ref{Table_Sim_pred} shows these results in terms of the square root of the mean-squared prediction errors (RMSPEs) for the BHM and the EHM, where the predictive mean of $y_t(\mb{s})$ ($y$-scale) or $p_t(\mb{s})\equiv g^{-1}(y_t(\mb{s}))$ ($p$-scale) is used. We refer to EHM-EM when the EM estimates of parameters are substituted in, and to EHM-TRUE when the true model parameters are substituted in. This latter case is expected to yield the best prediction and forecasting results.  

We observe that for the missing-by-design scenario, BHM gives a much smaller RMSPE value than EHM-EM. This indicates that accounting for the parameter variability improves predictions at locations without any observations nearby. For the missing-at-random scenario where nearby observations are available for those predictions, the prediction performances of BHM and EHM-EM are closer, although BHM still gives a smaller RMSPE value. It is not surprising that EHM-TRUE yields the smallest RMSPE for both scenarios, because it substitutes in the true parameter values that were used to simulate the data.  When predicting the underlying process on the probability scale, namely $\{p_t(\mb{s})\}$, the RMSPEs are much closer in comparison. Nevertheless, similar conclusions hold. The right panels of Figure~\ref{Fig_Sim_Pred_Y} show the predictive map of $\{y_t(\mb{s}_j^u)\}$ for the BHM, which looks very much like the map of their true values (in the middle panels).  

Compared with prediction at times when observations are available, forecasting the future is inherently more difficult, as illustrated by the larger RMSPE values in the last two columns of Table~\ref{Table_Sim_pred} compared to those in the first four columns. Nevertheless, on the $y$-scale, the same conclusions hold for forecasting as for prediction. The lower-right panel of Figure~\ref{Fig_Sim_Pred_Y} shows the forecasts for the BHM on the $y$-scale and, while the general pattern of ``high in the middle decreasing to low on the edges" is preserved, the shape of the signal is not. 

Next, we checked whether the inference of the proposed Bayesian hierarchical model is sensitive to a fixed value of $\sigma^2_{\xi}$. The EM estimate is $\hat{\sigma}^2_{\xi}=0.050$, which we propose as a ``plug-in" for Bayesian inference on all other unknowns. In our sensitivity study, we plug-in $\sigma^2_{\xi}\in\{0, 0.025, 0.050~(\textnormal{EM estimate}), 0.075, 0.100\}$, and then we use our methodology and computations to obtain the inference results under these values of $\sigma^2_{\xi}$. Table~\ref{Table_Sim_sensitivity} gives the parameter-estimation results of the scalar parameters in our model under different plug-in values of $\sigma^2_{\xi}$, except for $\sigma^2_{\xi}=0$. We can see that the inference results are very similar, and the $95\%$ credible intervals of the model parameters cover their respective true values. A reviewer suggested we try the extreme case $\sigma^2_{\xi}=0$ (i.e., no fine-scale-variation process in the model); however, this resulted in MCMC samples that clearly did not converge to a stationary distribution, and so that model was not supported by the data. We conclude that our inference for the Bayesian hierarchical model of the Arctic sea-ice-extent data is not very sensitive to the specification of the ``plug in" for the fine-scale-variation variance $\sigma^2_{\xi}$, and the hybrid inference remains valid. 

{\begin{table}[!ht]
\caption{Parameter-inference results for the BHM with $\sigma^2_{\xi}$ fixed at plug-in values of $\sigma^2_{\xi}$ given by $0.025, 0.050~(\textnormal{EM estimate}), 0.075,$ and $0.100$. The true values (in parentheses of the first column), posterior means, and the $95\%$ credible intervals of parameters (in parentheses) are reported. }
\label{Table_Sim_sensitivity}
\centering
\vspace{8pt}
\scalebox{0.65}{\begin{tabular}{|c|c|c|c|c|}
\toprule
Parameter& $\sigma^2_{\xi}=0.025$& $\sigma^2_{\xi}=0.050$ (EM) & $\sigma^2_{\xi}=0.075$& $\sigma^2_{\xi}=0.100$\\
\hline
$\beta_{0}~(5)$&$5.039~(4.910,5.168)$& $5.045~(4.923,5.169)$ &.$5.037~(4.911,5.160)$ & $5.042~(4.914,5.169)$\\
$\beta_1~(-15)$&$-15.140~(-15.492,-14.802)$&$-15.184~(-15.518,-14.851)$  &$-15.178~(-15.511,-14.846)$ & $-15.198~(-15.537,-14.851)$\\
$\lambda_1~(0.4)$&$0.236~(-0.087,0.548)$&$0.244~(-0.062,0.537)$  &$0.291~(0.005,0.614)$ & $0.257~(-0.044,0.548)$\\
$\lambda_2~(0.4)$ &$0.301~(0.073,0.530)$&$0.307~(0.078,0.528)$  &$0.340~(0.120,0.564)$ &  $0.321~(0.106,0.543)$\\
$\lambda_3~(0.035)$&$0.041~(-0.004,0.091)$& $0.046~(-0.004,0.097)$  &$0.038~(-0.009,0.085)$ & $0.040~(-0.005,0.084)$\\
\bottomrule
\end{tabular}}
\end{table}
}

\section{Spatio-Temporal Dynamics for Different Arctic Regions} \label{Sec_Append_Hov_local}

\begin{figure}[ht!]
\centering
  \includegraphics[width=11cm, height=7cm]{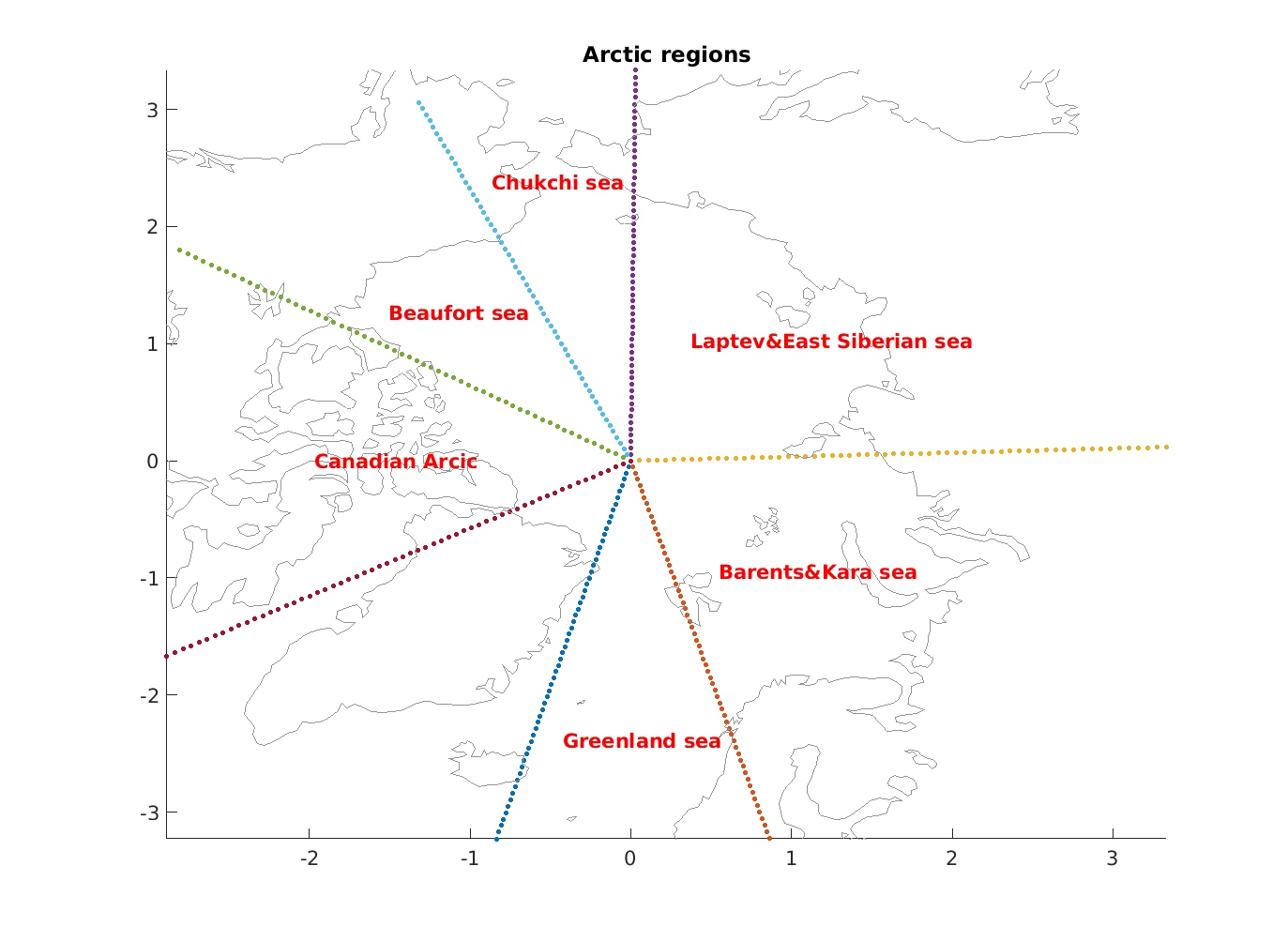}
\caption{The six different Arctic regions for sea ice, following the Arctic Regional Ocean Observing System website.}\label{Fig_RD_Arctic_regions}
\end{figure}

\begin{figure}[ht!]
\centering
\begin{subfigure}[Greenland sea]{
  \includegraphics[width=9cm, height=5cm]{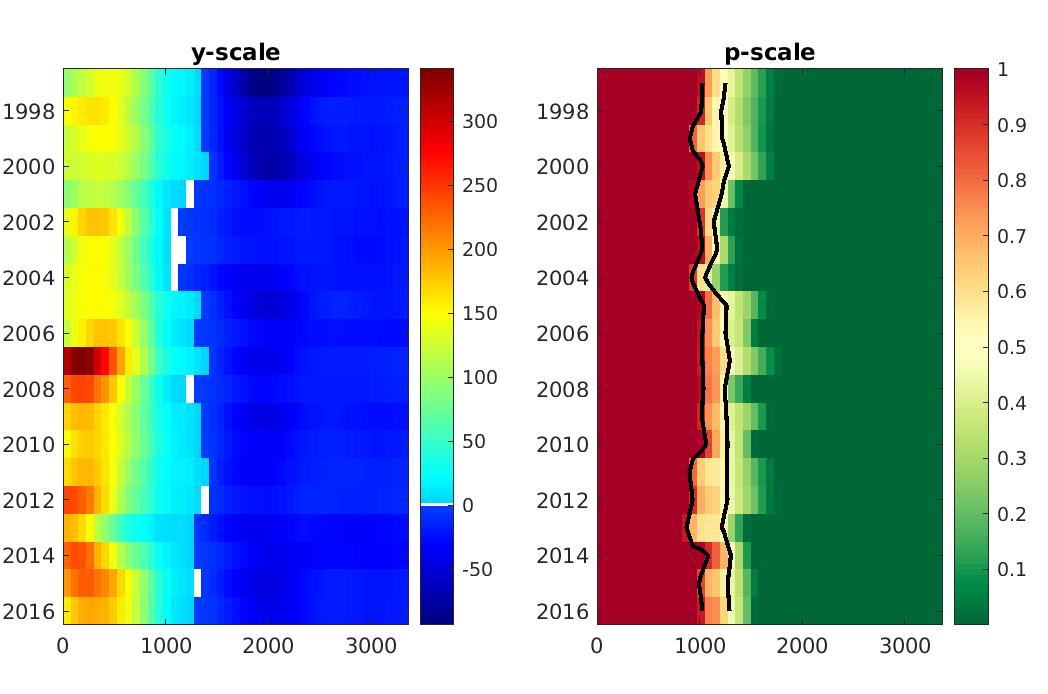}}
\end{subfigure}
\begin{subfigure}[Laptev\&East Siberian sea]{
  \includegraphics[width=10cm, height=5cm]{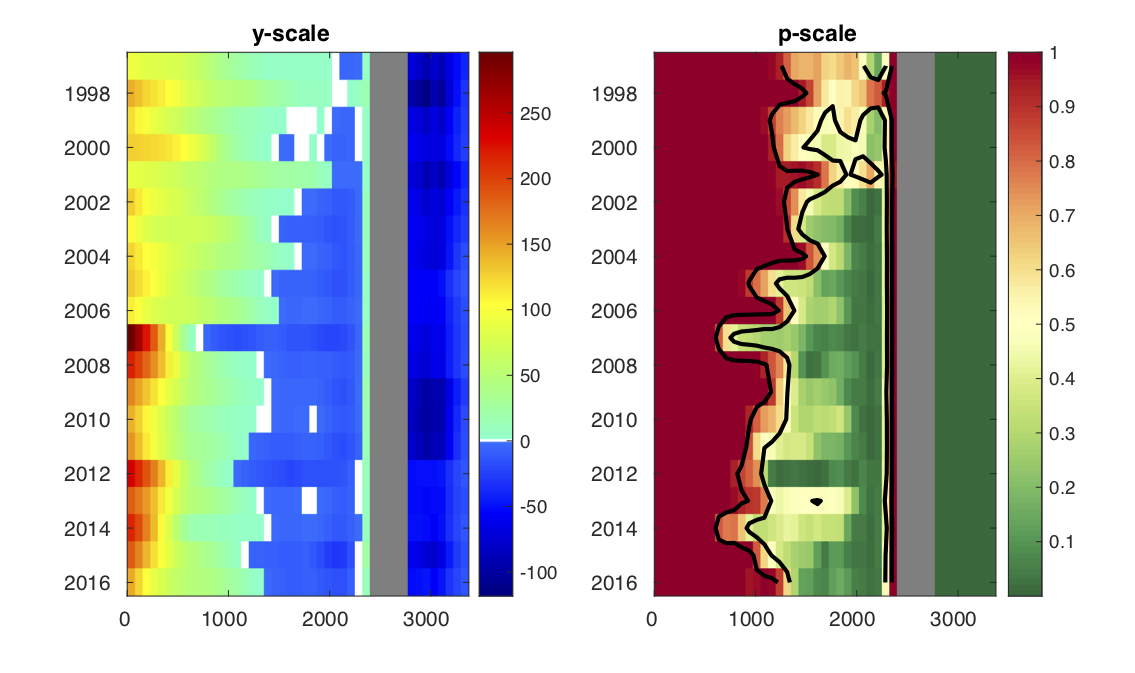}}
\end{subfigure}
\caption{Hovm\"{o}ller diagrams for the Greenland-sea region and the Laptev\&East-Siberian-sea region for both $\{y_t(\mb{s})\}$ and $\{p_t(\mb{s})\}$, where the grey strip indicates distances without observations. The image plots shown in the right panel are enhanced with contours of $\{p_t(\mb{s})\}$ for probabilities of $0.9$ and $0.5$ from left to right. The horizontal axis shows distance from the North Pole (in units of km) and the vertical axis shows time (in units of years). }\label{Fig_RD_Hov_local}
\end{figure}

In this section, we partitioned the entire Arctic into six regions and considered the temporal dynamics of the latent process for each of them. Following the website, Arctic Regional Ocean Observing System (\url{https://arctic-roos.org/observations/sea-ice-variability-in-regions}), we considered the six Arctic regions given by the map in Figure~\ref{Fig_RD_Arctic_regions}: The Greenland sea, the Barents\&Kara sea, the Laptev\&East Siberian sea, the Chukchi sea, the Beaufort sea, and the Canadian Arctic region. Note that the region north of Greenland is not considered, because it is always covered by sea ice. 

\begin{figure}[ht!]
\centering
\begin{subfigure}[Barents\& Kara sea]{
  \includegraphics[width=9cm, height=5cm]{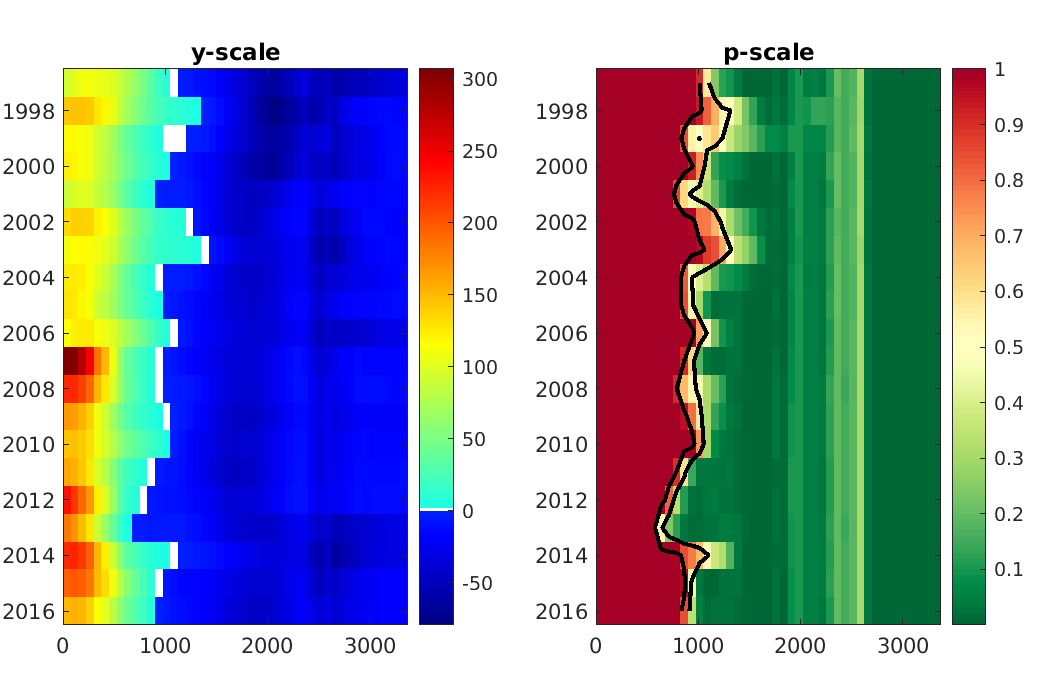}}
\end{subfigure}
\begin{subfigure}[Chukchi sea]{
  \includegraphics[width=9cm, height=5cm]{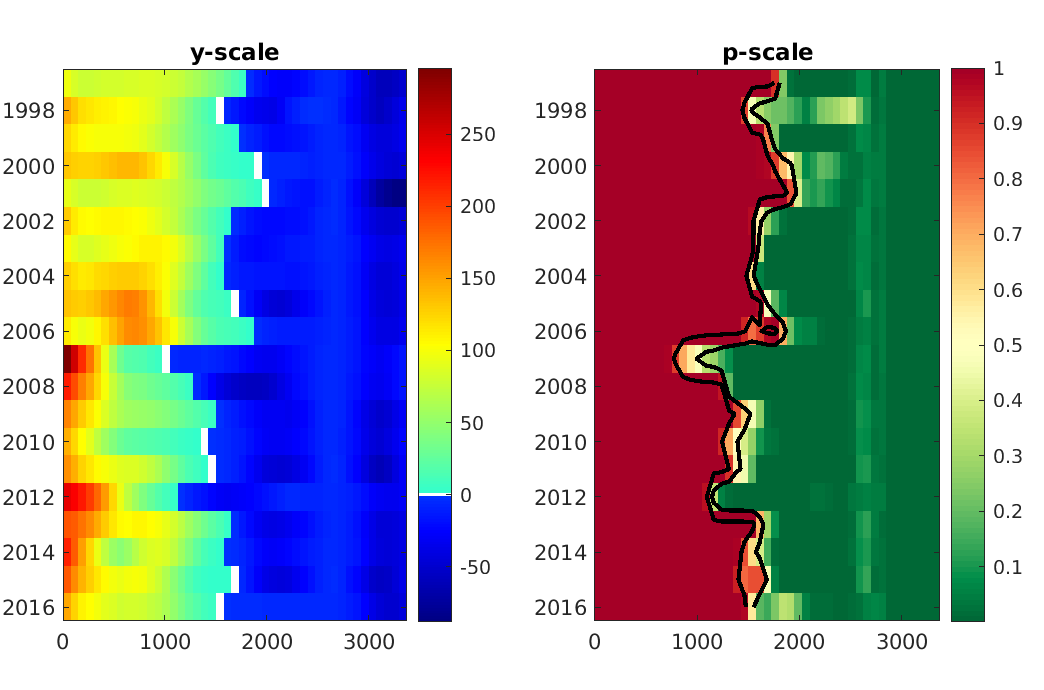}}
\end{subfigure}
\caption{Hovm\"{o}ller diagrams for the Barents\&Kara-sea region and the Chukchi-sea region for both $\{y_t(\mb{s})\}$ and $\{p_t(\mb{s})\}$. The image plots shown in the right panels are enhanced with contours of $\{p_t(\mb{s})\}$ for probabilities of $0.9$ and $0.5$ from left to right. The horizontal axis shows distance from the North Pole (in units of km) and the vertical axis shows time (in units of years). }\label{Fig_RD_Hov_local_more1}
\end{figure}

\begin{figure}[ht!]
\centering
\begin{subfigure}[Beaufort sea]{
\hspace{-0.7cm}  \includegraphics[width=9.4cm, height=5cm]{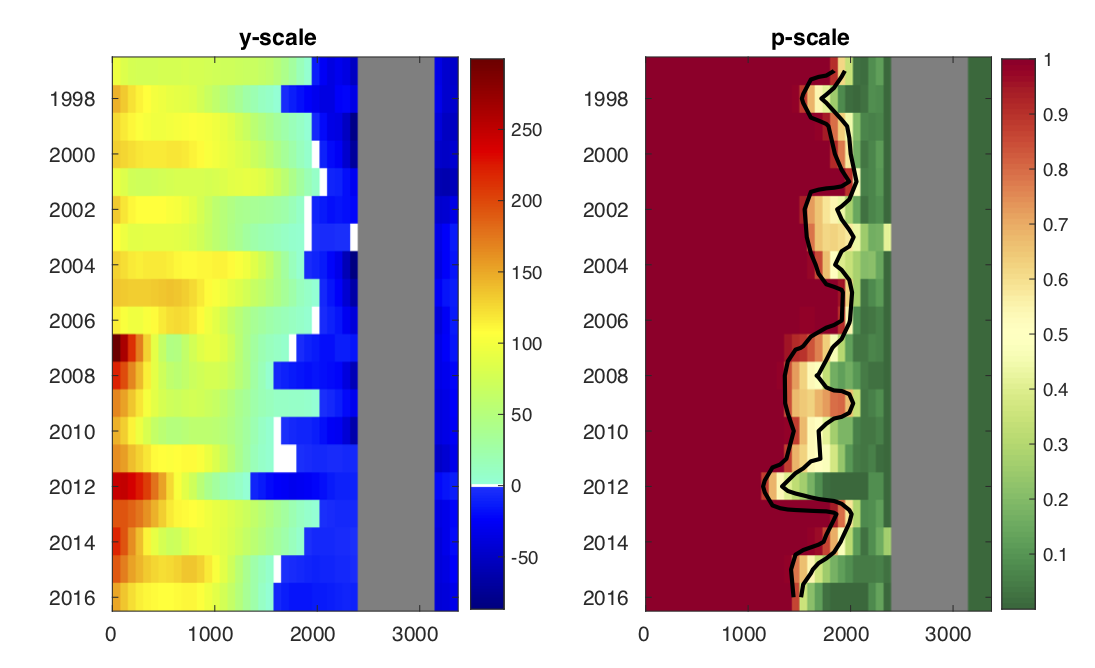}}
\end{subfigure}
\begin{subfigure}[Canadian Arctic region]{
  \includegraphics[width=9cm, height=5cm]{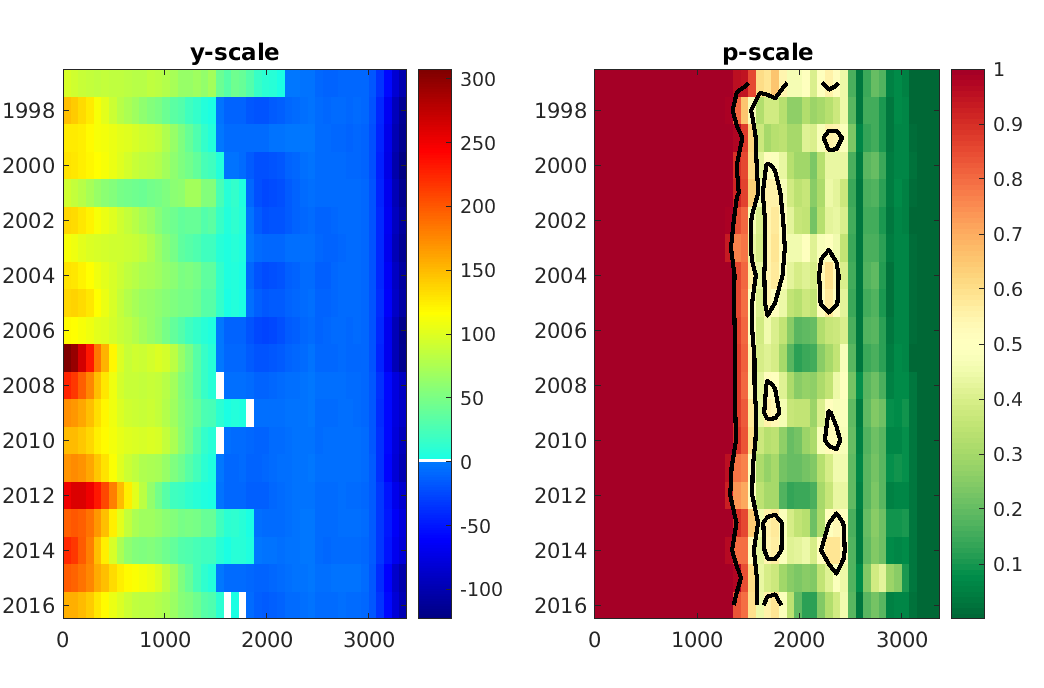}}
\end{subfigure}
\caption{Hovm\"{o}ller diagrams for the Beaufort-sea region and the Canadian-Arctic region for both $\{y_t(\mb{s})\}$ and $\{p_t(\mb{s})\}$, where the grey strip indicates distances without observations.. The image plots shown in the right panels are enhanced with contours of $\{p_t(\mb{s})\}$ for probabilities of $0.9$ and $0.5$ from left to right. The horizontal axis shows distance from the North Pole (in units of km) and the vertical axis shows time (in units of years). }\label{Fig_RD_Hov_local_more2}
\end{figure}

In a similar manner to the summaries for the entire Arctic in Section~4.2, we obtained Hovm\"{o}ller diagrams for the six regions, which are shown in Figures~\ref{Fig_RD_Hov_local}\textendash\ref{Fig_RD_Hov_local_more2}. Figure~\ref{Fig_RD_Hov_local} shows Hovm\"{o}ller diagrams for two Arctic regions, the Greenland-sea region and the Laptev\&East-Siberian-sea region. It can be observed that for the Greenland sea, which is an open sea connected to the Atlantic ocean, the temporal changes of the latent process $\{p_t(\mb{s})\}$ are very small, implying that sea ice is not very sensitive to climate forcing for this region. In contrast, for the combined region of the Laptev\&East Siberian sea, the latent process $\{p_t(\mb{s})\}$ is much more variable over time, and its contours clearly move northwards in certain years. 

Figures~\ref{Fig_RD_Hov_local_more1} and \ref{Fig_RD_Hov_local_more2} show Hovm\"{o}ller diagrams for the remaining four regions. We found that the temporal variability of $\{p_t(\mb{s})\}$ at each latitude band is very small for the Canadian-Arctic region, indicating a relatively stable sea-ice cover there. In contrast, the other three Arctic regions show larger temporal variabilities, and the contours of $\{p_t(\mb{s})\}$ clearly move further north in 2007 and 2012, when the Arctic SIE decreased substantially. For example, in 2012 the Arctic SIE reached its minimum, and the water\textendash ice contours of $\{p_t(\mb{s})\}$ for the Chukchi sea and the Beaufort sea clearly moved northwards, indicating substantial loss of sea ice in those regions. In comparison, the water\textendash ice contours of $\{p_t(\mb{s})\}$ for the Barents\&Kara sea oscillated over time, but only showed a move northwards in 2013.  

In summary, when looking at the temporal dynamics of $\{p_t(\mb{s})\}$ regionally, different Arctic regions show different patterns of variability and some regions seem to be more sensitive to climate forcing than others. More scientific attention should be paid to the Laptev\&East-Siberian-sea region, the Chukchi-sea region, and the Beaufort-sea region, where larger temporal variabilities of the latent process are observed. 

\section{Sampling details and convergence diagnostics}\label{Sec_algorithm_diagnostics}

\subsection{Initialization}

In this subsection, we describe specification of the parameters in the prior distributions of $\mb{K}$ and $\mb{U}$, and the proposal distributions of $\sbf{\beta}$, $\{\lambda_j\}$ (in the propagator matrix $\mb{H}$), $\{\sbf{\eta}_t\}$, and $\{\sbf{\xi}_t\}$. These parameters are specified using their respective EM estimates \citep{zhang2018estimating}. Specifically, for $\pi(\mb{K})\sim \textnormal{IW}(\nu_K,\sbf{\Phi}_K)$ and $\pi(\mb{U})\sim \textnormal{IW}(\nu_U, \sbf{\Phi}_U)$, we set $\nu_{K}=\nu_{U}=2r$ in order to make $\{\sbf{\eta}_t: t=1,\ldots, T\}$ well constrained in the prior when modeling the Arctic sea-ice-extent data; then $\sbf{\Phi}_{K}=(3r+1)\hat{\mb{K}}$ and $\sbf{\Phi}_{U}=(3r+1)\hat{\mb{U}}$ to make the priors of $\mb{K}$ and $\mb{U}$ concentrate around their respective EM estimates, denoted as $\hat{\mb{K}}$ and $\hat{\mb{U}}$. For $\{\lambda_j\}\in (-1, 1)$, we reparameterized them as $\{\lambda_j\equiv\frac{e^{\tau_{\lambda_j}}-1}{e^{\tau_{\lambda_j}}+1}\}$ such that $\{\tau_{\lambda_j}\}\in \mathbb{R}$; then, the proposal distribution of $\tau_{\lambda_j}$ was specified to be $\textnormal{Gau}(\hat{\tau}_{\lambda_j},a_{\lambda_j}V_{\lambda_j})$, where $\hat{\tau}_{\lambda_j}$ is the EM estimate of $\tau_{\lambda_j}$, $V_{\lambda_j}$ is the estimated posterior variance of $\tau_{\lambda_j}$ obtained from the Hessian matrix evaluated at the estimated posterior modes, and $a_{\lambda_j}$ is the step size to be adjusted to achieve a proper acceptance rate. In a similar manner, we specified the Gaussian proposal distributions for $\sbf{\beta}$, $\{\sbf{\eta}_t\}$, and $\{\sbf{\xi}_t\}$.

\subsection{MCMC sampling algorithm}

\begin{algorithm}[ht!]
\caption{\hspace{-3pt}: MCMC algorithm} \label{Algorithm_MCMC}
\begin{algorithmic}
\\ 1. Initialization: Specify $\pi(\mb{K})$ and $\pi(\mb{U})$, and the proposal distributions for $\sbf{\beta}$, $\{\lambda_j\}$, $\{\sbf{\eta}_t\}$, and $\{\sbf{\xi}_t\}$.
\\ 2. At $k=0$, select the starting values $\{\sbf{\eta}^{(k)}, \sbf{\xi}^{(k)}, \sbf{\theta}^{(k)}\}$.
\\3. At $(k+1)$-th iteration,\For{$t=1$ to $T$}
\\\hspace{1cm}Simulate from the full conditional distribution $p(\sbf{\eta}_t^{(k+1)}|\sbf{\xi}^{(k)}, \sbf{\theta}^{(k)})$.
\EndFor
\\4. \For{$t=1$ to $T$}
\For{$i=1$ to $N_t$}\\
\hspace{1cm}Simulate from the full conditional distribution $p(\xi_{t,i}^{(k+1)}|\sbf{\eta}^{(k+1)}, \sbf{\theta}^{(k)})$.
\EndFor
\EndFor
\\5. Simulate from the full conditional distributions for $\mb{K}$ and $\mb{U}$, and simulate using the adaptive Metropolis-Hastings algorithm for $\{\sbf{\beta}, \lambda_1, \lambda_2, \lambda_3\}$.
\\6. Repeat steps 3~\textendash~5 until the target number of samples $\{\sbf{\eta}^{(k)}, \sbf{\xi}^{(k)}, \sbf{\theta}^{(k)}\}$ is attained.
\\5. Discard an initial number of samples, which is the ``burn-in" period.
\end{algorithmic}
\end{algorithm} 

Our computations were done on a multi-core machine, which allowed parallel computation. For the random effects $\{\sbf{\eta}_t\}$, we sampled them blockwise; that is, we sampled the random vectors $\{\sbf{\eta}_t\}$ one-by-one. Each $\sbf{\eta}_t$ has a relatively low dimension of $217$, which can be easily sampled. The MCMC samples of $\{\sbf{\eta}_t\}$ are indeed correlated, so to improve the mixing of the MCMC chain, we applied the adaptive MCMC Algorithm~5 in \cite{andrieu2008tutorial}, which iteratively adjusts the covariance matrix of the proposal distribution. The proposal distribution for generating the $(k+1)$-th posterior sample of $\sbf{\eta}_t$ , denoted by $\sbf{\eta}_t^{(k+1)}$, is $\textnormal{Gau}(\sbf{\eta}_t^{(k)}, a_1\mb{V}_{\sbf{\eta},t})$. Following \cite{sengupta2016predictive}, the covariance matrix $\mb{V}_{\sbf{\eta},t}$ is obtained by the same EM algorithm given in \cite{zhang2018estimating}, which is the estimated posterior covariance matrix of $\sbf{\eta}_t$ obtained based on the Hessian matrix evaluated at the posterior modes. We adjusted the step size $a_1$ to achieve an acceptance rate between $26\%$ to $50\%$ for each $\sbf{\eta}_t$, $t=1,2,\ldots, T$. Similarly, we sampled each $\xi_{t,i}^o$ individually in parallel, using the same adaptive MCMC algorithm referred to above \citep{andrieu2008tutorial}. The summary of the MCMC algorithm is given in Algorithm~\ref{Algorithm_MCMC}.

\subsection{Convergence diagnostics}
\begin{figure}[ht!]
\centering
  \includegraphics[width=10cm, height=6cm]{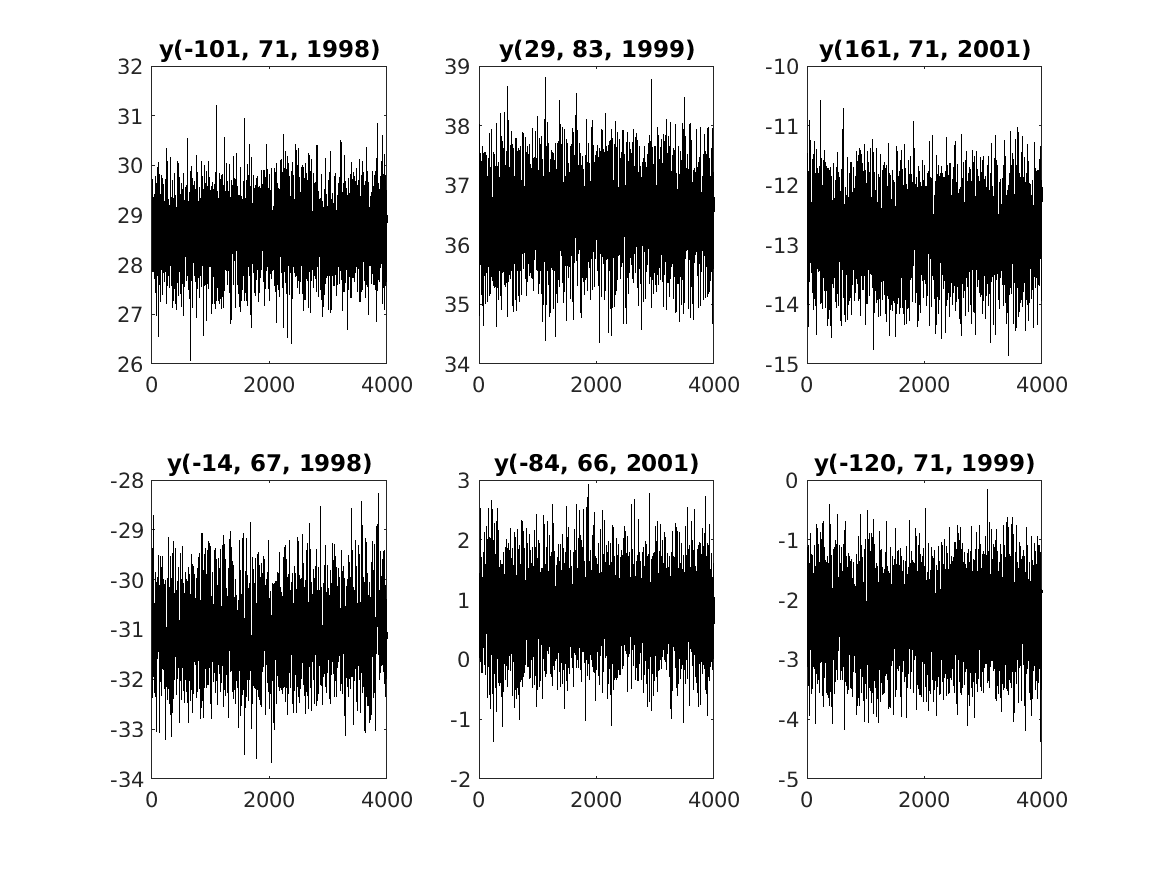}
\caption{The trace plots of the predictive samples of $\{y_t(\mb{s})\equiv y(\textnormal{lon}, \textnormal{lat}, t)\}$, where six pairs of $\mb{s}\equiv (\textnormal{lon}, \textnormal{lat})$ and $t\in\{1997, \ldots, 2001\}$ were randomly selected. }\label{Fig_RD_TP_Y}
\end{figure}

\begin{figure}[ht!]
\centering
\begin{subfigure}[Regression coefficients, $\{\beta_j\}$]{
  \includegraphics[width=12cm, height=8cm]{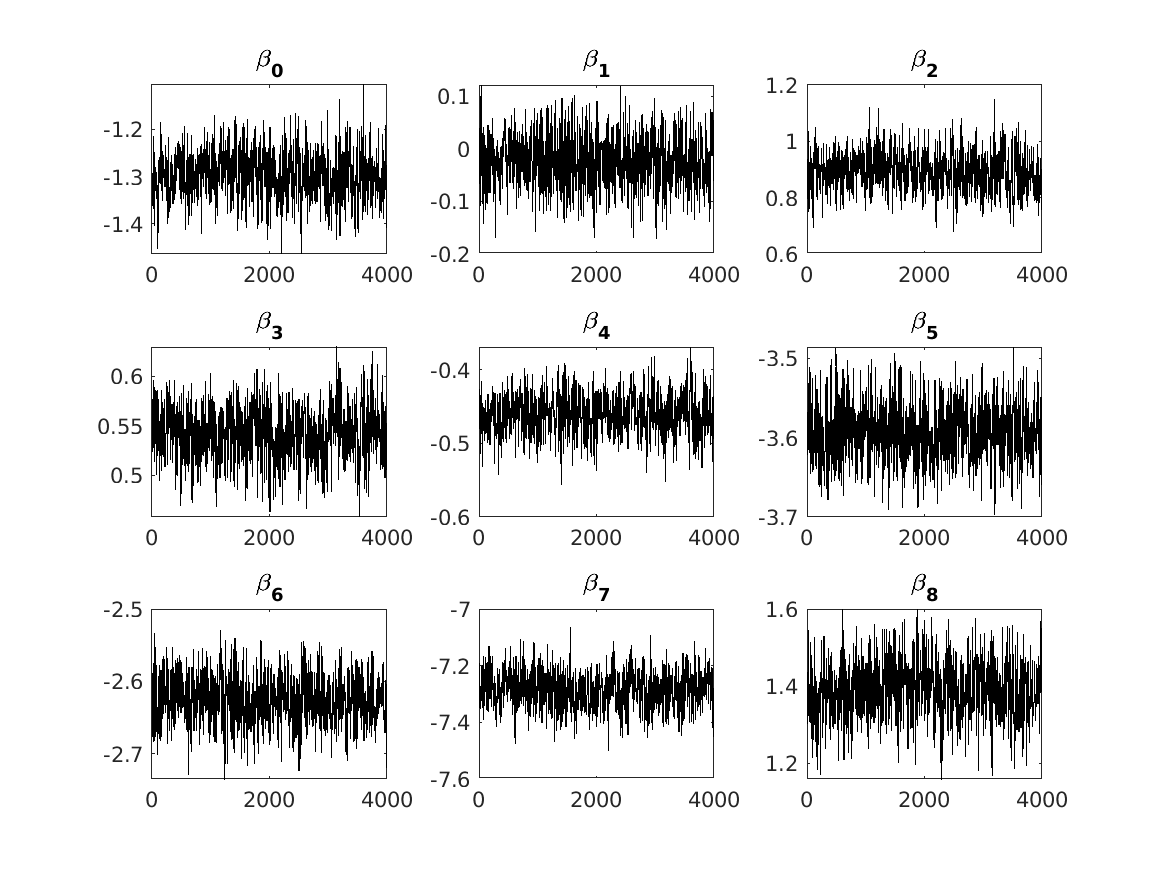}}
\end{subfigure}
\begin{subfigure}[Propagator-matrix parameters, $\{\lambda_j\}$]{
  \includegraphics[width=10cm, height=6cm]{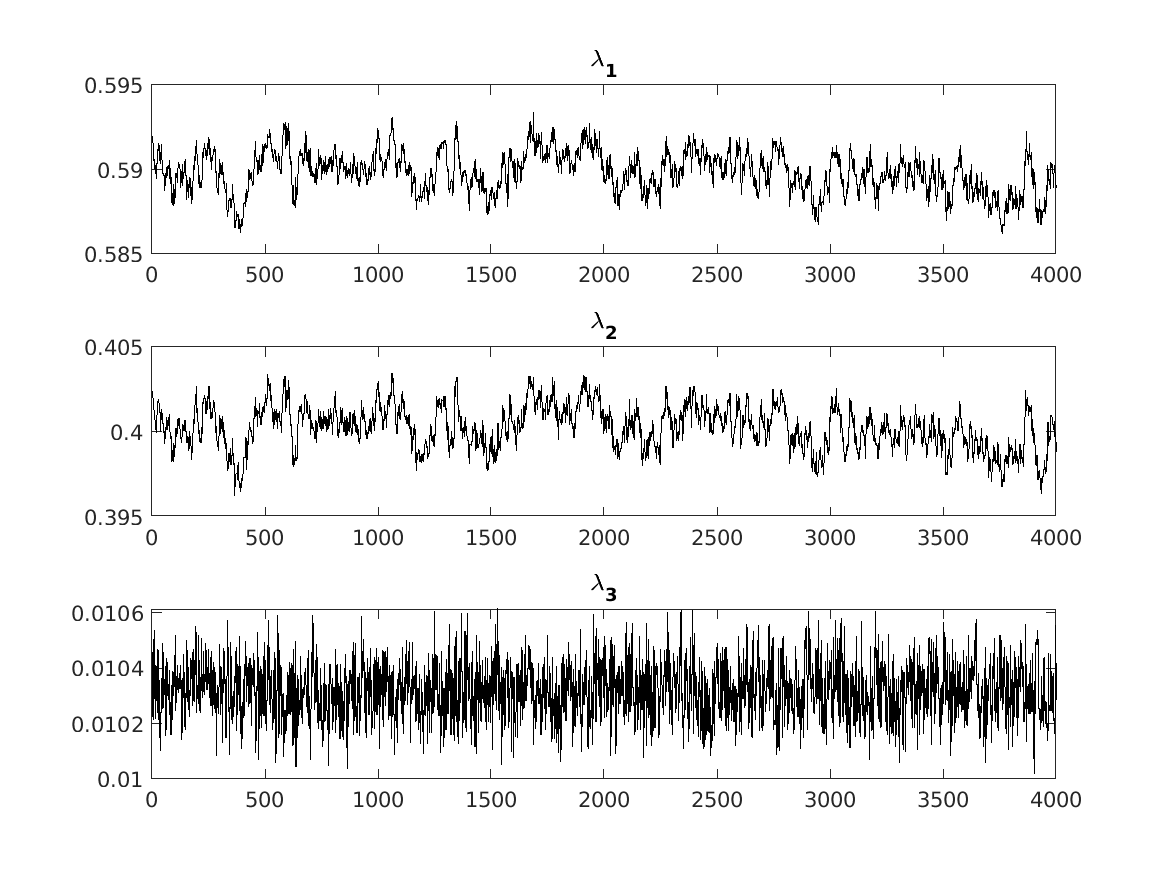}}
\end{subfigure}
\caption{The trace plots of the MCMC samples of the regression coefficients, $\{\beta_j\}$ (upper panel), and the correlation parameters in the propagator matrix, $\mb{H}$ (lower panel) for Period~1.}\label{Fig_RD_TP_mean_lambda}
\end{figure}

\begin{figure}[ht!]
\centering
  \includegraphics[width=12cm, height=8cm]{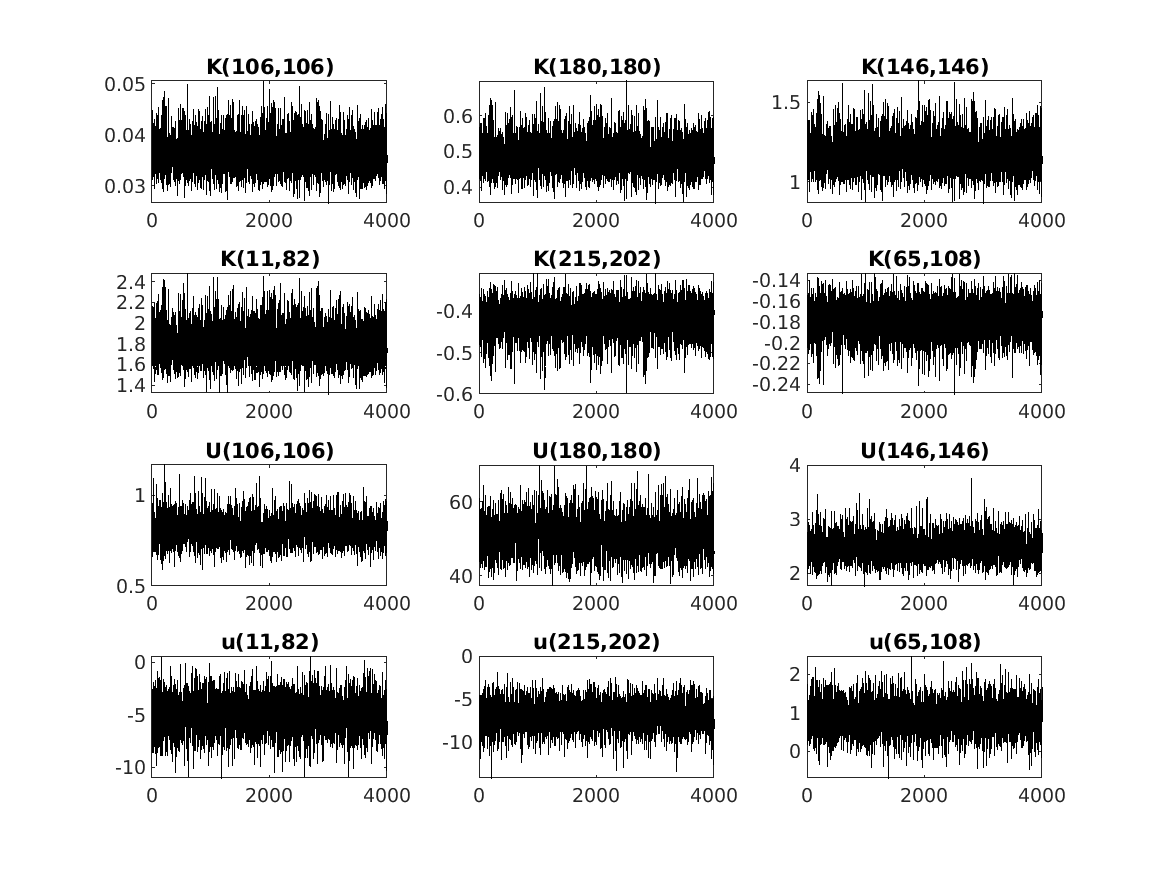}
\caption{The trace plots of the MCMC samples of selected entries of the covariance matrices, $\mb{K}$ and $\mb{U}$. Three diagonal entries and three off-diagonal entries were randomly selected for $\mb{K}$ and $\mb{U}$. }\label{Fig_RD_TP_K_U}
\end{figure}

For the Arctic sea-ice-extent data, we ran the MCMC chains for $14,000$ iterations with the first $2,000$ samples discarded as the burn-in period. We collected $4,000$ posterior samples after burn-in by thinning the chain and retaining the samples from every third iteration. We take the results in Period 1 as an example. The emphasis in this research is on prediction: Figure~\ref{Fig_RD_TP_Y} shows the trace plots of the MCMC chains of six randomly selected values of the latent process $\{y_t(\mb{s})\}$, and they illustrate that the chains are sampling from their corresponding stationary distributions. Figures~\ref{Fig_RD_TP_mean_lambda} and \ref{Fig_RD_TP_K_U} show the trace plots of the MCMC samples for selected model parameters in the Bayesian hierarchical model. The MCMC chains of regression coefficients and the entries of $\mb{K}$ and $\mb{U}$ mixed well and converged quickly, while the MCMC samples of $\lambda_1$ and $\lambda_2$ in the propagator matrix $\mb{H}$ appear to have relatively large correlations and are slow to converge.

Now, recall that the focus of our paper is to propose useful functionals on the predictive distribution of $\{y_t(\mb{s})\}$ that provide different ``views" of the changes of Arctic sea ice. The quick convergence of the predictive samples of $\{y_t(\mb{s})\}$ guarantees that the subsequent analyses (such as the averaged predictive means of $\{y_t(\mb{s})\}$ and $\{p_t(\mb{s})\}$, the empirical semivariograms, the Arctic-region Hovm\"{o}ller diagrams, and the dynamic plots of the ice-to-water and water-to-ice transitions in Section~S5) are based on appropriate predictive distributions. 

\section{Classification accuracy for Arctic sea-ice-extent data} \label{Sec_classification}

\begin{figure}[ht!]
\centering
\begin{subfigure}[$t=2001$]{
  \includegraphics[width=8cm, height=3.5cm]{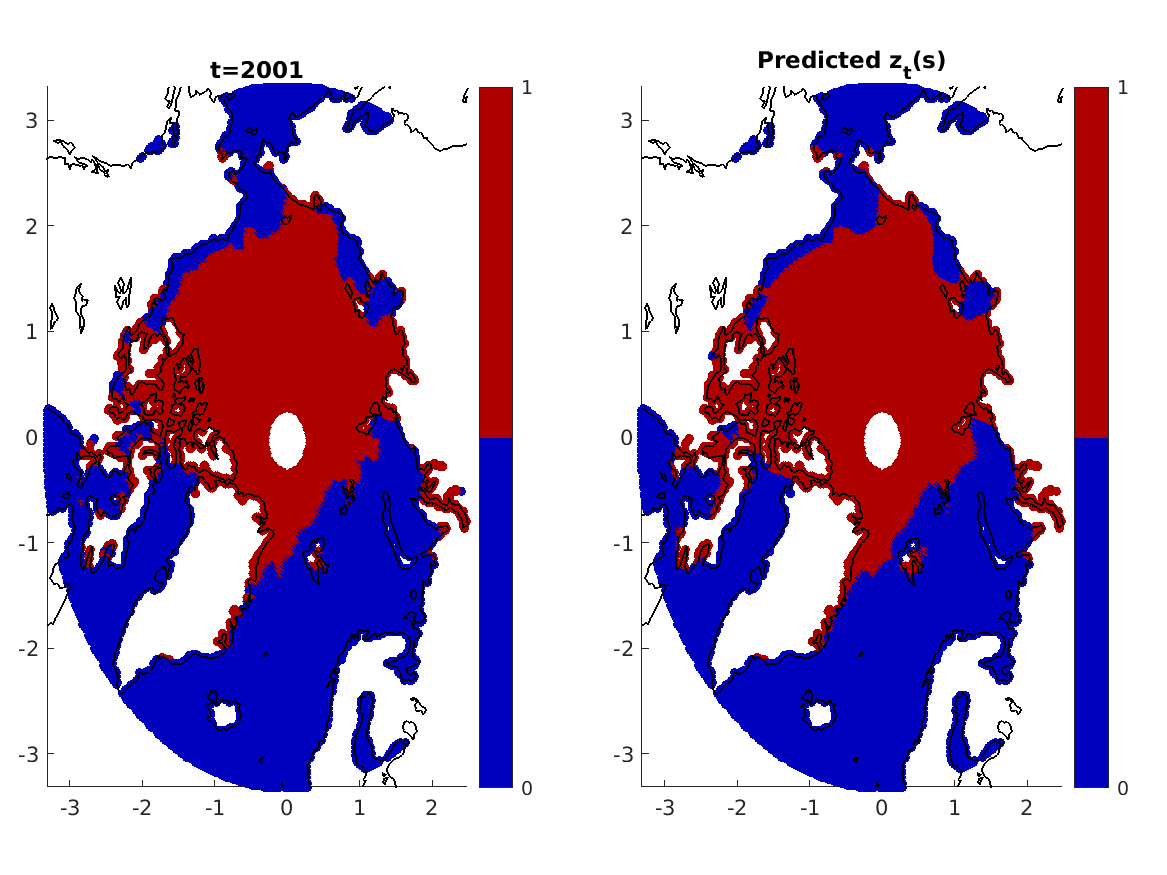}}
\end{subfigure}
\begin{subfigure}[$t=2016$]{
  \includegraphics[width=8cm, height=3.5cm]{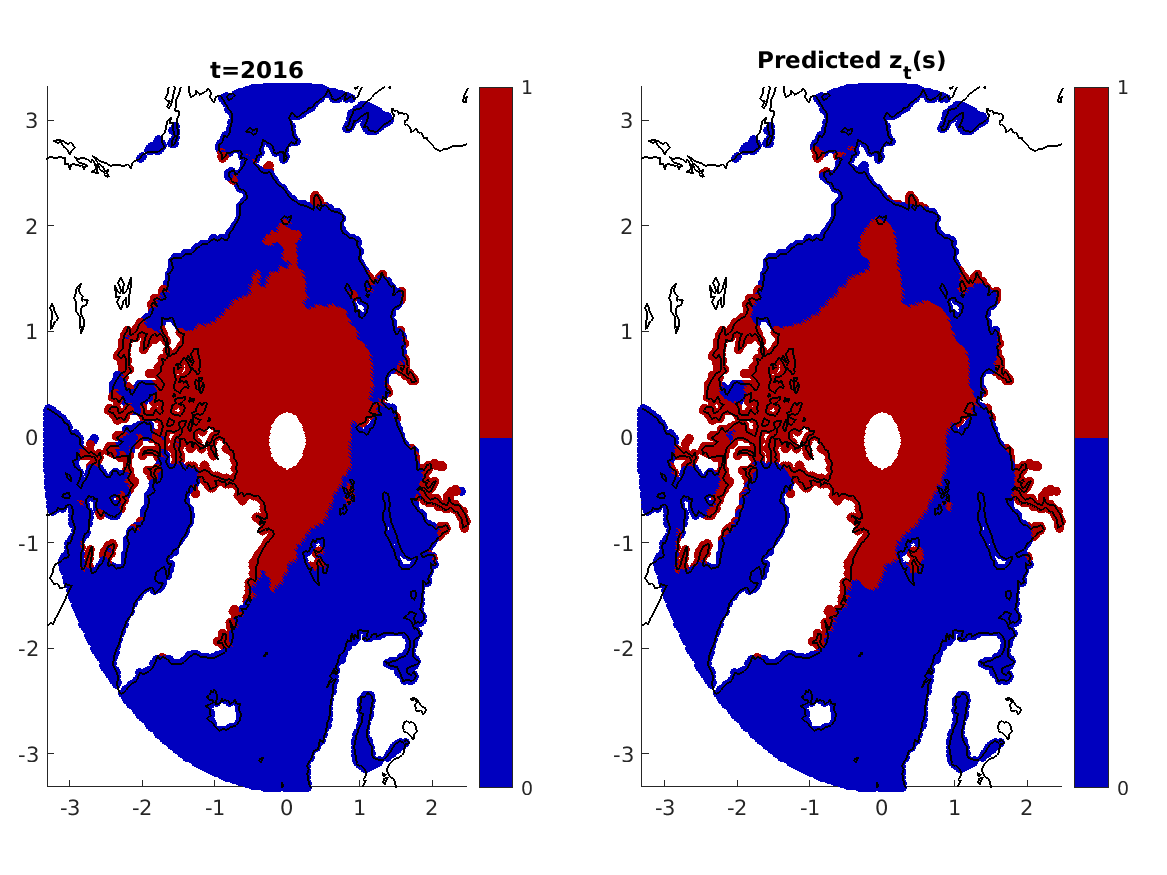}}
\end{subfigure}
\caption{The observed data (left panels) versus the predicted binary surface $I(E(p_t(\mb{s})|\textnormal{data})\leq 0.15)$ (right panels) for $t=2001$ and $t=2016$. }\label{Fig_RD_classification}
\end{figure}

We use the posterior mean of $p_t(\mb{s})$ and a cut-off value of $0.15$ to give a binary classification and compare it to the data; see Table~\ref{Table_cl_rate}. Our fitted model leads to a classification accuracy greater than or equal to $95\%$ for each year. Figure~\ref{Fig_RD_classification} shows the observed data and the predicted binary values $I(E(p_t(\mb{s})|\textnormal{data})\leq 0.15)$ for years $t=2001$ and $t=2016$. We can see that the misclassifications typically appear in the ice\textendash water boundaries (e.g., in the Canadian-Archipelago region where ice pixels are surrounded by water pixels). Overall, the predicted binary surfaces match the observed ones very well. 
\begin{table}[!ht]
\caption{The classification accuracy per year using the posterior mean of $p_t(\mb{s})$ and the cut-off value of $0.15$. } \label{Table_cl_rate}
\label{Table_RD_PE}
\centering
\vspace{8pt}
\scalebox{1.0}{\begin{tabular}{|c|c|c|c|c|c|}
\toprule
Period 1 & 1997 & 1998 & 1999& 2000& 2001\\
Accuracy& $97\%$ &    $96\%$&   $95\%$&  $96\%$ &  $97\%$\\
\hline
Period 2 & 2002 & 2003 & 2004& 2005& 2006\\
Accuracy& $96\%$ &  $97\%$ &  $98\%$& $97\%$& $96\%$\\
\hline
Period 3 & 2007 & 2008 & 2009& 2010& 2011\\
Accuracy& $96\%$ &  $95\%$ & $96\%$ & $95\%$ & $95\%$\\
\hline
Period 4 & 2012 & 2013 & 2014& 2015& 2016\\
Accuracy& $96\%$ & $97\%$ & $97\%$ & $96\%$ & $96\%$\\
\hline
\bottomrule
\end{tabular}}
\end{table}

\section{Visualization of the ice-to-water and water-to-ice transition probabilities} \label{Sec_IWT_WIT}


\begin{figure}[ht!]
\begin{center}
\begin{subfigure}[$t+1=2001$]{
\includegraphics[width=9cm, height=4cm]{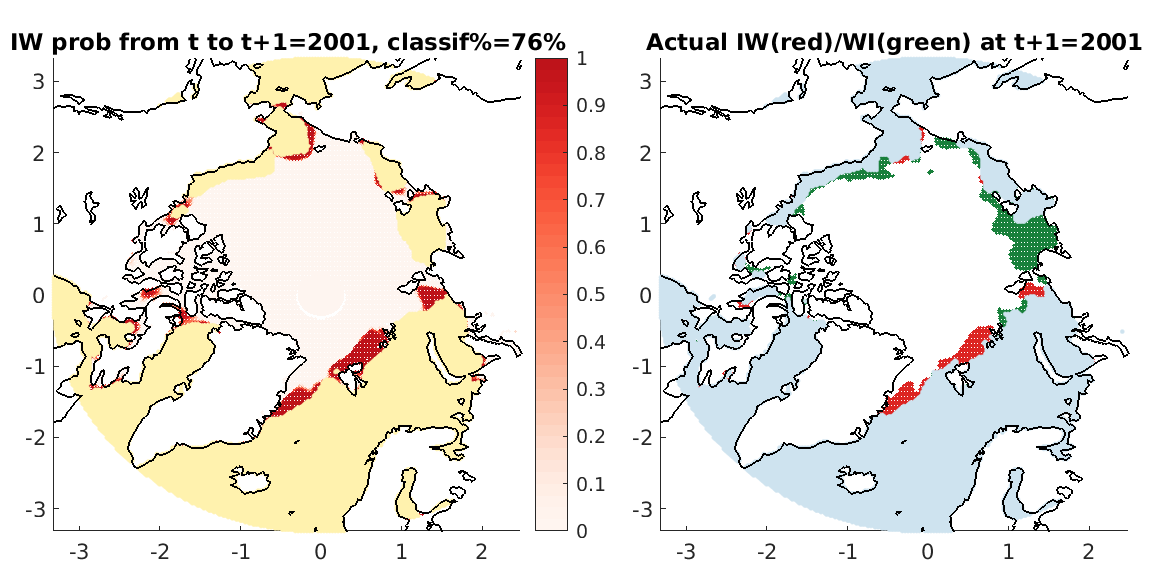}}
\end{subfigure}
\begin{subfigure}[$t+1=2016$]{
\includegraphics[width=9cm, height=4cm]{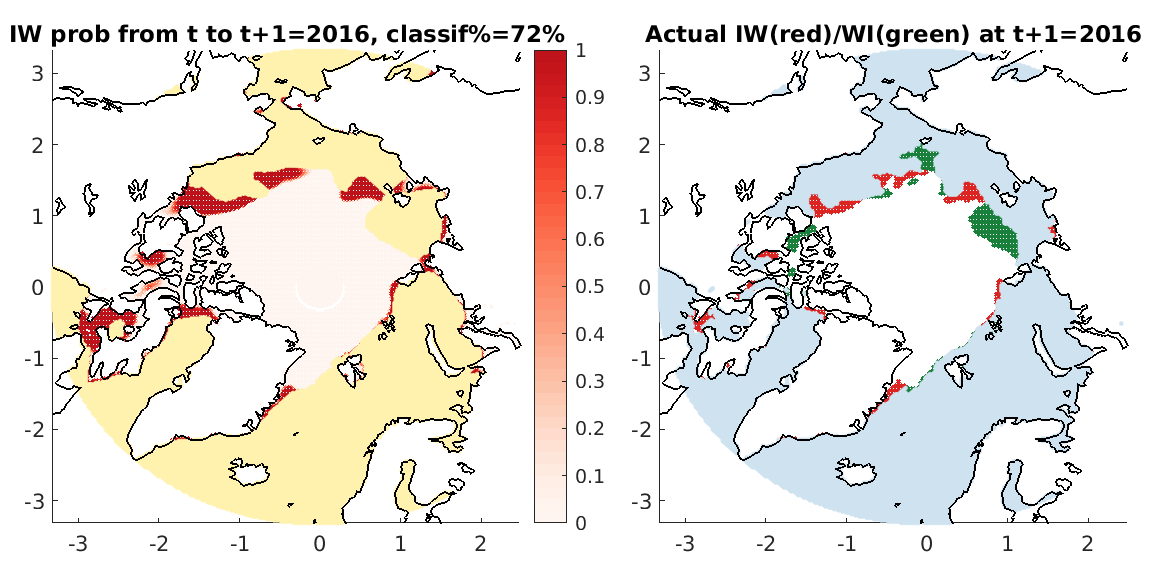}}
\end{subfigure}
\caption{The spatial maps of ice-to-water transition probabilities (left panel) and the observed loss (red) and gain (green) of ice (right panel) at $t+1=2001$ and $t+1=2016$. Pixels colored yellow in left panels have $P(p_t(\mb{s})\ge 0.15|\textnormal{data})=0$ (i.e., are water pixels at time $t$). Pixels colored blue in right panels are water pixels at both time $t$ and time $t+1$.}\label{Anime_IWT}
\end{center}
\end{figure}

Following \cite{zhang2018estimating}, the ``risk" that a spatial pixel at location $\mb{s}$ is in water at time $t+1$, \textit{given} it is in ice at time $t$, is the \textit{conditional probability}, $\pi_{t|t+1}^{\textnormal{IW}}(\mb{s})\equiv P(p_t(\mb{s})\ge0.15, p_{t+1}(\mb{s})<0.15|\mb{Z}^o)/P(p_t(\mb{s})\ge0.15|\mb{Z}^o)$. This conditional probability is the \textit{ice-to-water transition probability} at pixel $\mb{s}$ and time $t+1$. Spatial regions where this risk is high are of most concern. Figure~\ref{Anime_IWT} shows the spatial maps of these ice-to-water transition probabilities (left panel) at $t+1=2001$ and $t+1=2016$, where high values (shown in darker red) indicate areas with retreating sea ice; the yellow background represents (water) pixel $\mb{s}$ where $P(p_t(\mb{s})\ge0.15|\mb{Z}^{o})=0$ (so that $\pi_{t|t+1}^{\textnormal{IW}}(\mb{s})$ is undefined). The observed loss and gain of ice from time $t$ to $t+1$ is shown (right panel) in Figure~\ref{Anime_IWT}; the red represents previous ice pixels at time $t$ that transition to water pixels at time $t+1$, and the green represents previous water pixels at time $t$ that transition to ice pixels at time $t+1$. The IW classification rate at year $t+1$ is defined as
\begin{eqnarray*}
R^{\textnormal{IW}}_{t+1}=\frac{\sum\limits_{i=1}^{N}I(\pi_{t|t+1}^{\textnormal{IW}}(\mb{s}_i)>0.5)\cdot I(z_t(\mb{s}_i)\ge0.15, z_{t+1}(\mb{s}_i)<0.15)}{\sum\limits_{i=1}^{N}I(z_t(\mb{s}_i)\ge0.15, z_{t+1}(\mb{s}_i)<0.15)}\,,
\end{eqnarray*}
where $I(\cdot)$ is the indicator function, the denominator gives the number of pixels that were ice at time $t$ and water at time $t+1$ (i.e., the observed ice-to-water transition pixels), and the numerator gives the number of such pixels that are correctly identified by using a $0.5$ cut-off for the ice-to-water transition probability.

\begin{figure}[ht!]
\begin{center}
\begin{subfigure}[$t+1=2001$]{
\includegraphics[width=9cm, height=4cm]{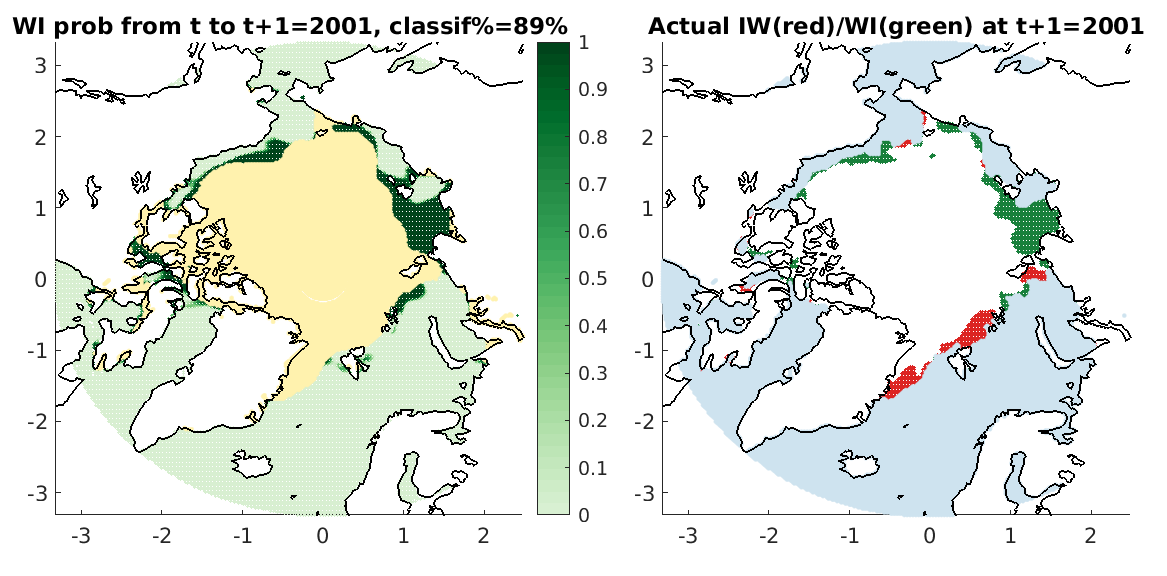}}
\end{subfigure}
\begin{subfigure}[$t+1=2016$]{
\includegraphics[width=9cm, height=4cm]{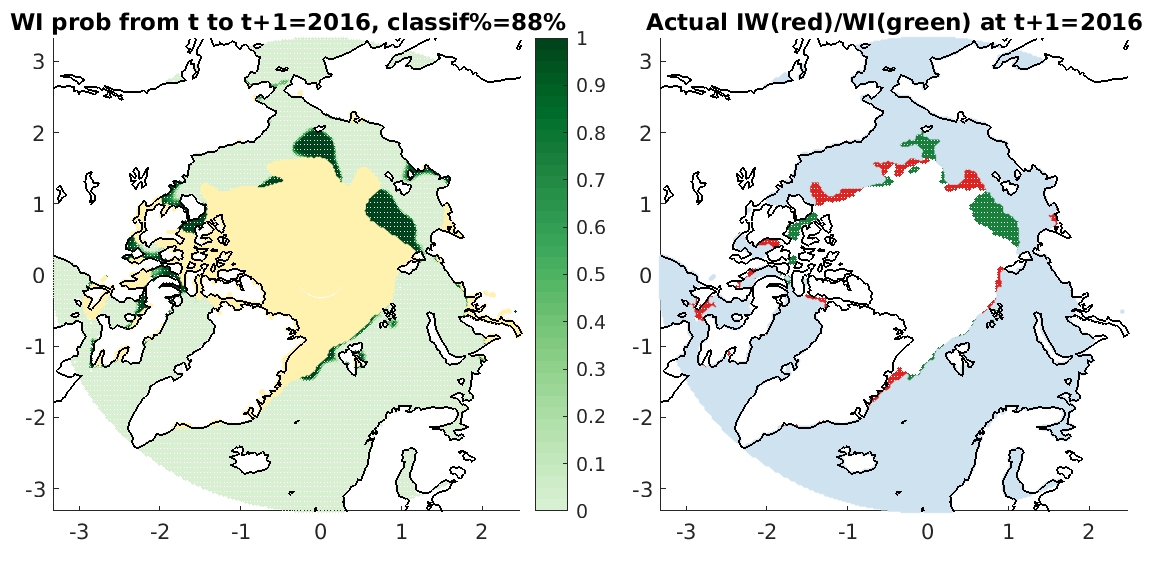}}
\end{subfigure}
\caption{The spatial maps of water-to-ice transition probabilities (left panel) and the observed loss (red) and gain (green) of ice (right panel) at $t+1=2001$ and $t+1=2016$. Pixels colored yellow in left panels have $P(p_t(\mb{s})< 0.15|\textnormal{data})=0$ (i.e., are ice pixels at time $t$). Pixels colored blue in right panels are water pixels at both time $t$ and time $t+1$.}\label{Anime_WIT}
\end{center}
\end{figure}

Similarly, the conditional probability, $\pi_{t|t+1}^{\textnormal{WI}}(\mb{s})\equiv P(p_t(\mb{s})<0.15, p_{t+1}(\mb{s})\ge0.15|\mb{Z}^o)/P(p_t(\mb{s})<0.15|\mb{Z}^o)$, gives the \textit{water-to-ice transition probability} at pixel $\mb{s}$ and time $t+1$. Figure~\ref{Anime_WIT} shows the spatial maps of the water-to-ice transition probabilities at $t+1=2001$ and $t+1=2016$, where high values (shown in darker green) indicate the areas that are likely to gain ice in the following year, and the yellow background represents (ice) pixel $\mb{s}$ where $P(p_t(\mb{s})<0.15|\mb{Z}^{o})=0$ (so that $\pi_{t|t+1}^{\textnormal{WI}}(\mb{s})$ is undefined). The right panel of Figure~\ref{Anime_WIT} is identical to that given in Figure~\ref{Anime_IWT}; that is, the red represents previous ice pixels at time $t$ that transition to water pixels at time $t+1$, and the green represents previous water pixels at time $t$ that transition to ice pixels at time $t+1$. The WI classification rate at year $t+1$ is defined as
\begin{eqnarray*}
R^{\textnormal{WI}}_{t+1}=\frac{\sum\limits_{i=1}^{N}I(\pi_{t|t+1}^{\textnormal{WI}}(\mb{s}_i)>0.5)\cdot I(z_t(\mb{s}_i)<0.15, z_{t+1}(\mb{s}_i)\ge 0.15)}{\sum\limits_{i=1}^{N}I(z_t(\mb{s}_i)<0.15, z_{t+1}(\mb{s}_i)\ge 0.15)}\,,
\end{eqnarray*}
where $I(\cdot)$ is the indicator function, the denominator gives the number of pixels that were water at time $t$ and ice at time $t+1$ (i.e., the observed water-to-ice transition pixels), and the numerator gives the number of such pixels that are correctly identified by using a $0.5$ cut-off for the water-to-ice transition probability.


\bibliographystyle{chicago}
\bibliography{ref-ST-GLM-supp}